%% file: paper.tex
\documentclass[aps,prapplied,reprint]{revtex4-1} %
\usepackage[utf8x]{inputenc}
\usepackage{graphicx}
\usepackage{amsmath}
\usepackage{amssymb}
\usepackage{wasysym}
\usepackage{tabularx}
\usepackage{tablestyles}
\input{eigendefs}

\usepackage{etoolbox}

\newtoggle{submissionBuild}
\settoggle{submissionBuild}{true}

\nottoggle{submissionBuild}{%
  \usepackage[noheader]{gitver}
  \usepackage{soul}
  \sethlcolor{green}
}{}

\makeatletter
\appto\abstract{%
  \let\latexlist\list \rightskip=\leftskip
  \def\list{\edef\keeprightskip{\the\rightskip}\latexlist}%
  \patchcmd\latexlist{\ignorespaces}{\rightskip\keeprightskip\ignorespaces}{}{}%
}
\makeatother

\begin{document}
\title{A robust, high-flux source of laser-cooled ytterbium atoms}

\author{E. Wodey}
\author{R. J. Rengelink}
\author{C. Meiners}
\author{E. M. Rasel}
\author{D. Schlippert}
\affiliation{Leibniz Universität Hannover, Institut für Quantenoptik, Welfengarten 1, 30167 Hannover, Germany}

\iftoggle{submissionBuild}{%
  \date{\today{}}
}{%
  \date{\today{} -- \hl{\mbox{\#\gitVer}}}%
}

\begin{abstract}
  We present a high-flux source of cold ytterbium atoms that is robust, lightweight and low-maintenance. Our apparatus delivers \SI{1e9}{\atoms\per\second} into a 3D magneto-optical trap without requiring water-cooling or high current power supplies. We achieve this by employing a Zeeman slower and a 2D magneto-optical trap fully based on permanent magnets in Halbach configurations. This strategy minimizes mechanical complexity, stray magnetic fields, and heat production while requiring little to no maintenance, making it applicable to both embedded systems that seek to minimize electrical power consumption, and large scale experiments to reduce the complexity of their subsystems.
\end{abstract}

\maketitle

\section{Introduction}

\input{sec-introduction}

\section{Oven with microchannel nozzle}
\label{sec:oven}

\input{sec-oven}

\section{Permanent magnet Zeeman slower in Halbach configuration}
\label{sec:zeeman-slower}

\input{sec-zs}

\section{2D-MOT recollimation/deflection stage}
\label{sec:2d-mot}

\input{sec-2d-mot}

\section{Full system characterization}
\label{sec:3d-mot}

\input{sec-3d-mot}

\section{Conclusion}
\label{sec:conclusion}

\input{sec-conclusion}

\section*{Acknowledgements}

This work is part of the Hannover very long baseline atom interferometry facility, a major research equipment funded by the German Research Foundation (Deutsche Forschungsgemeinschaft, DFG). We acknowledge support by the Collaborative Research Centers 1128 ``geo-Q'' and 1227 ``DQ-mat'', and Germany's Excellence Strategy within EXC-2123 ``QuantumFrontiers'' (project No. 390837967). D.~S. acknowledges funding from the German Federal Ministry of Education and Research (BMBF) through the funding program Photonics Research Germany (contract No. 13N14875). E.~W. acknowledges support from ``Nieders\"achsisches Vorab'' through the ``Quantum- and Nano-Metrology~(QUANOMET)'' initiative (project No. QT3). We thank M.~Robert-de-Saint-Vincent for insightful discussions on the microchannel nozzle and in-vacuum mirrors, and D.~Tell for careful proof-reading of the manuscript. We are grateful to C.~Schubert, D.~Tell, and K.~Zipfel for their contributions and W.~Ertmer for his vision and long-lasting support towards very long baseline atom interferometry in Hannover.

\bibliographystyle{apsrev4-2}
\bibliography{paper}

\end{document}

%% file: eigendefs.tex
\usepackage[per-mode=repeated-symbol,detect-all,forbid-literal-units]{siunitx}
\usepackage{xspace}
\usepackage{nicefrac}
\usepackage{tikz}
\usepackage{xcolor}
\usepackage{chemformula}

\usetikzlibrary{plotmarks}
\usetikzlibrary{arrows}

\newcommand{\Yb}[1]{\ensuremath{^{#1}\text{Yb}}\xspace}

\newcommand{\Toven}{\ensuremath{T_{\mathrm{oven}}}\xspace}

\newcommand{\Ntubes}{\ensuremath{N_{\mathrm{tubes}}}\xspace}
\newcommand{\term}[3]{\ensuremath{{}^{#1}{#2}_{#3}}\xspace}
\renewcommand{\term}[3]{\ch{^{#1}{#2}_{#3}}\xspace}
\newcommand{\singletterms}{\ensuremath{\term{1}{S}{0}\to\term{1}{P}{1}}\xspace}
\newcommand{\tripletterms}{\ensuremath{\term{1}{S}{0}\to\term{3}{P}{1}}\xspace}
\newcommand{\Isat}{\ensuremath{I_{\mathrm{sat}}}\xspace}
\newcommand{\thetahalf}{\ensuremath{\theta_{1/2}}\xspace}
\newcommand{\thetahalftilde}{\ensuremath{\tilde{\theta}_{1/2}}\xspace}
\newcommand{\thetamax}{\ensuremath{\theta_{\mathrm{max}}}\xspace}
\newcommand{\amax}{\ensuremath{a_{\mathrm{max}}}\xspace}
\newcommand{\Lstop}{\ensuremath{L_{\mathrm{stop}}}\xspace}
\newcommand{\unitv}[1]{\ensuremath{\hat{e}_{#1}}\xspace}
\newcommand{\Pmot}{\ensuremath{P_{\mathrm{MOT}}}\xspace}

\definecolor{cud-black}         {RGB}{0,0,0}
\definecolor{cud-orange}        {RGB}{230,159,0}
\definecolor{cud-sky-blue}      {RGB}{86,180,233}
\definecolor{cud-bluish-green}  {RGB}{0,158,115}
\definecolor{cud-yellow}        {RGB}{240,228,66}
\definecolor{cud-blue}          {RGB}{0,114,178}
\definecolor{cud-vermillion}    {RGB}{213,94,0}
\definecolor{cud-reddish-purple}{RGB}{204,121,167}

\DeclareRobustCommand\showmark[2]{\tikz[]{\node[#2]{\pgfuseplotmark{#1}};}}
\DeclareRobustCommand\showline[2]{\tikz[baseline=-0.5ex]{\draw[#1,line width=#2] (0,0) -- (0.3,0);}}

\DeclareSIUnit{\atom}{at}
\DeclareSIUnit{\atoms}{atoms}

%% file: sec-introduction.tex
Neutral cold atoms are a versatile resource for many domains of modern physics. They lie at the core of modern quantum engineering, realizing quantum simulators \cite{Gross2017} and proposals for quantum computers \cite{Saffman2016}. They also constitute a key element of state-of-the-art quantum metrology, producing the most stable frequency references \cite{Ludlow2015}, contributing to the determination of fundamental constants \cite{Rosi2014,Parker2018}, searching for new physics \cite{Safronova2018}, and enabling inertial sensing with unprecented stability \cite{Dutta2016,Freier2016}. Furthermore, owing to increased robustness and compactness of cold-atoms technology \cite{Hill2016,Grotti2018,Carraz2009,Grosse2016}, even very large scale experiments \cite{Canuel2018,Coleman2018,Zhan2019} and eminently challenging environments are accessible, including zero-g aircrafts \cite{Barrett2016}, sounding rockets \cite{Becker2018}, and orbiting spacecrafts \cite{Liu2018}. These developments are paving the way for long-term operation in space \cite{Frye2019} and a new era for quantum metrology \cite{Aguilera2014,Gao2018,Kolkowitz2016,Hogan2016}.

Light-pulse atom interferometry \cite{Hogan2008} is instrumental in many of these applications. It relies on the precise control over the phase of matter waves and is able to exploit the simultaneous interrogation of a large number of atoms to deliver accurate and stable measurements. Since each measurement destroys the atomic sample, minimizing aliasing and dead-time effects while preserving low noise characteristics for applications in metrology \cite{Savoie2018} and inertial navigation \cite{Canuel2006,Garrido2019} requires reliable high-flux sources of cold atoms. In addition, setups of all extents, but especially large-scale and transportable devices benefit tremendously from low maintenance, robust, low power consumption strategies for the initial atomic cooling steps in order to increase up-time and overall reliability.

Alkali atoms, mainly rubidium and cesium, are used to achieve the current state of the art in atom interferometry. Recently, however, ytterbium and other alkaline-earth-like atoms gained interest from the community due to their low magnetic susceptibility in the ground state and richer electronic structure. In particular, the availability of long-lived metastable states allows for novel coherent matterwave manipulation schemes \cite{Hu2017,Plotkin-Swing2018,Rudolph2020} and has implications for gravitational wave detection \cite{Loriani2019}. In addition, these elements constitute interesting choices for tests of the universality of free fall \cite{Hartwig2015}.
 
Alkaline-earth-like atoms have far lower vapor pressures compared to alkali atoms. This requires hot ovens to obtain sufficient content in the gaseous phase to initiate laser cooling. The correspondingly large atomic velocities at the exit of the oven prompted the use of Zeeman slowers \cite{Takasu2003} or, more recently, 2D magneto-optical traps \cite{Dorscher2013} to enable efficient loading into a three-dimensional trap. In their conventional implementation, however, these techniques are power consuming and usually lead to bulky and heavy setups as they feature high-power electrical circuits which require water cooling and associated maintenance.

\begin{figure}
  \centering
  \includegraphics[width=\linewidth]{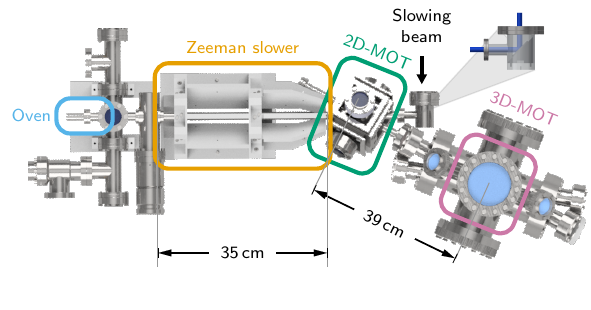}
  \caption{Overview of our cold ytterbium apparatus. An atomic beam is produced from metallic ytterbium chunks heated in an oven with a microchannel nozzle. The combination of a Zeeman slower and 2D-MOT slows down, redirects, and recollimates the atomic beam for delivery in a 3D-MOT.}
  \label{fig:setup}
\end{figure}

We present a robust, lightweight and low-maintenance source of slow ytterbium atoms delivering \SI{1e9}{\atom\per\second} initial loading rate into a three-dimensional magneto-optical trap (3D-MOT), comparable to other state-of-the-art strontium \cite{Yang2015} and ytterbium \cite{Lee2015} systems. A distinctive feature of our setup is the use of permanent magnets to generate all the required magnetic fields, thus reducing the weight and complexity of the setup while also requiring less electrical power and no water-cooling for operation.

Our apparatus is depicted in figure~\ref{fig:setup}. We describe and characterize its four distinct elements in the following sections. First, a hot atomic beam is produced from metallic ytterbium chunks in an oven terminated by a microchannel nozzle (sec.~\ref{sec:oven}). The mean atomic velocity in the beam is then reduced from around \SI{300}{\meter\per\second} to \SI{20}{\meter\per\second} in a Zeeman slower made of permanent magnets in a Halbach configuration (sec.~\ref{sec:zeeman-slower}). A 2D-MOT (sec.~\ref{sec:2d-mot}) serves as a deflection and recollimation stage, preventing the residual fast atoms from entering the main experimental chamber without introducing in-vacuum moveable parts, thus limiting failure points. Finally, the slowed atoms are captured in a 3D-MOT, demonstrating the usability and performance of our apparatus (sec.~\ref{sec:3d-mot}).

%% file: sec-oven.tex
\subsection{Design}

\begin{figure}
  \centering
  \includegraphics[width=\linewidth]{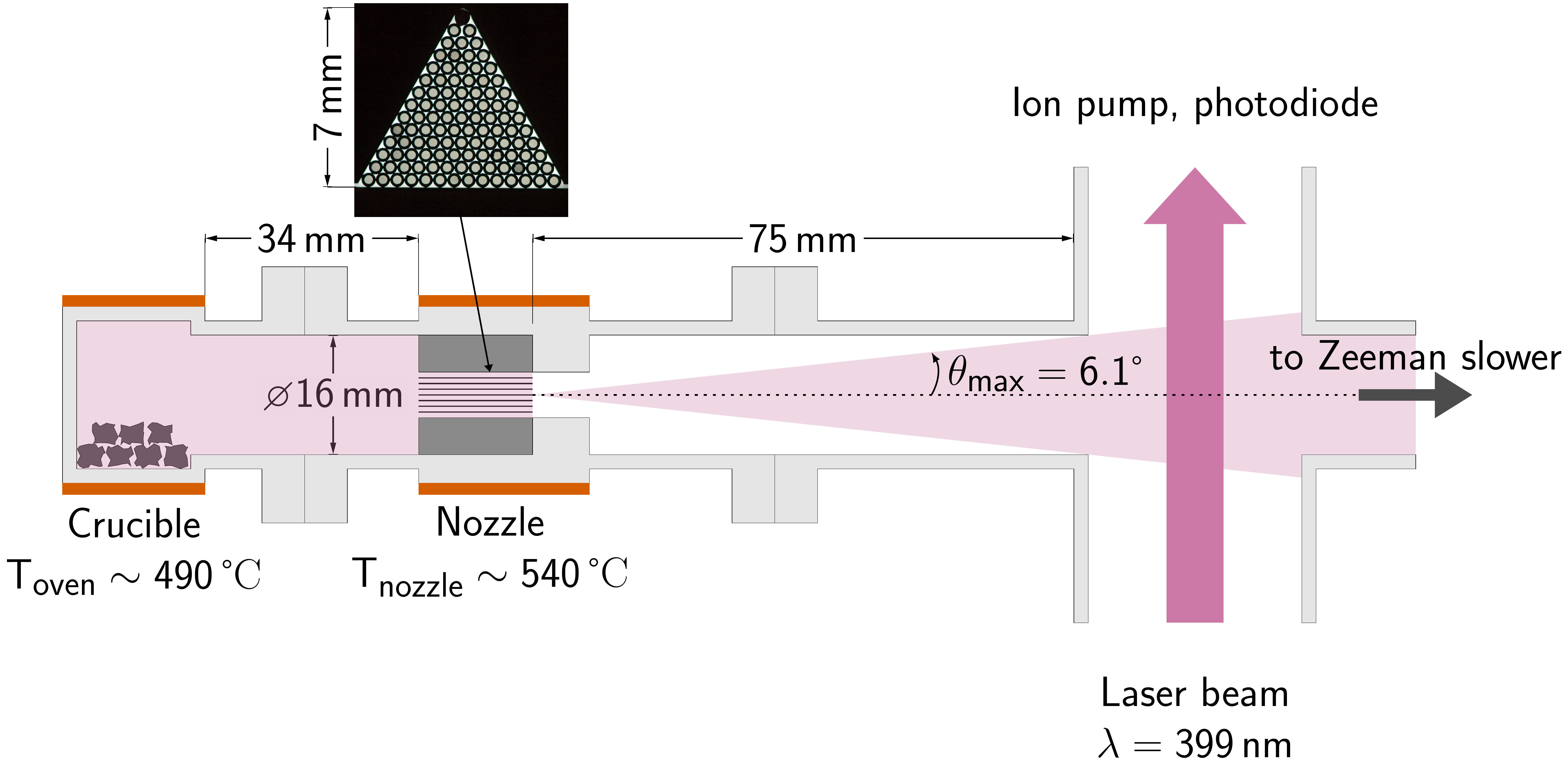}
  \caption{Sketch of the ytterbium beam apparatus. Solid ytterbium chunks are heated to $\Toven = \SI{490}{\celsius}$ to form partial ytterbium vapor at \SI{2e-2}{\milli\bar} using circular band heaters (depicted in red). A directed beam is extracted by a nozzle made of \num{104} large length-to-diameter ratio tubes (inset: photograph of the microchannel stackup with homogeneous backlight). The properties of the atomic beam (flux, divergence) are measured around \SI{100}{\milli\meter} downstream the nozzle by means of laser absorption spectroscopy near the \singletterms resonance at \SI{399}{\nano\meter}.}
  \label{fig:oven-setup}
\end{figure}

Figure~\ref{fig:oven-setup} shows the design of our ytterbium oven. About \SI{5}{\gram} of solid metallic chunks (natural isotopic abundance, approximate chunk volume \SI{10}{\milli\meter\cubed}) are vaporized in a cylindrical vacuum chamber with \SI{19}{\milli\meter} inner diameter. This crucible is connected to the oven's nozzle piece via CF (ConFlat) flanges. The vacuum annealed oxygen-free copper gasket sealing the connection is silver-plated to avoid corrosion by the hot ytterbium vapor. The nozzle is made of $\Ntubes = \num{104}$ microchannels assembled in a quasi-hexagonal lattice constrained in a triangular holder (inset of fig.~\ref{fig:oven-setup}). The microchannels are AISI 316L stainless-steel capillaries of inner diameter $2a = \SI{280}{\micro\meter}$ and outer diameter \SI{320}{\micro\meter} cut to length $L = \SI{12}{\milli\meter}$. The triangular prism holder allows for an almost defect-free lattice, minimizing apertures larger than the microchannel diameter $2a$ \cite{Senaratne2015}. The microchannels are stacked inside the holder and clamped from one face of the prism. However, machining the prism's edge opposing the clamping face with conventional milling techniques leads to a non-negligible finite radius of curvature. To properly constrain the microchannel array, we insert a $\diameter{}\SI{0.5}{\milli\meter}$ tungsten wire in place of the capillary along this edge of the nozzle array (top of inset in figure~\ref{fig:oven-setup}).

We use mineral insulated, $\SI{25.4}{\milli\meter}$ diameter band heaters with integrated K-type thermocouples to heat the crucible and nozzle piece and monitor their temperature. The thermocouples are read out using Maxim Inc. MAX31850K 1-wire digitizing integrated circuits. Isolating the whole assembly with mineral whool and aluminum foil, we use only \SI{55}{\watt} of electrical power to maintain the crucible at \SI{490}{\celsius} and the nozzle piece around \SI{40}{\celsius} higher to avoid clogging of the microchannels. A major loss channel is thermal conduction through the vacuum chamber. We measured temperatures up to \SI{100}{\celsius} around \SI{100}{\milli\meter} downstream of the nozzle part.

The atomic beam extracted from the vapor through the nozzle travels towards the Zeeman slower in a vacuum of approximately \SI{1e-8}{\milli\bar} maintained by a \SI{30}{\liter\per\second} ion pump. Before entering the Zeeman slower, the beam traverses an area offering a \SI{38}{\milli\meter}-diameter optical access on an axis perpendicular to the atomic beam. This enables the use of laser absorption spectroscopy to characterize the flux and divergence of the atomic beam.

\subsection{Ytterbium atomic beam}

\begin{figure}
  \centering
  \includegraphics[width=\linewidth]{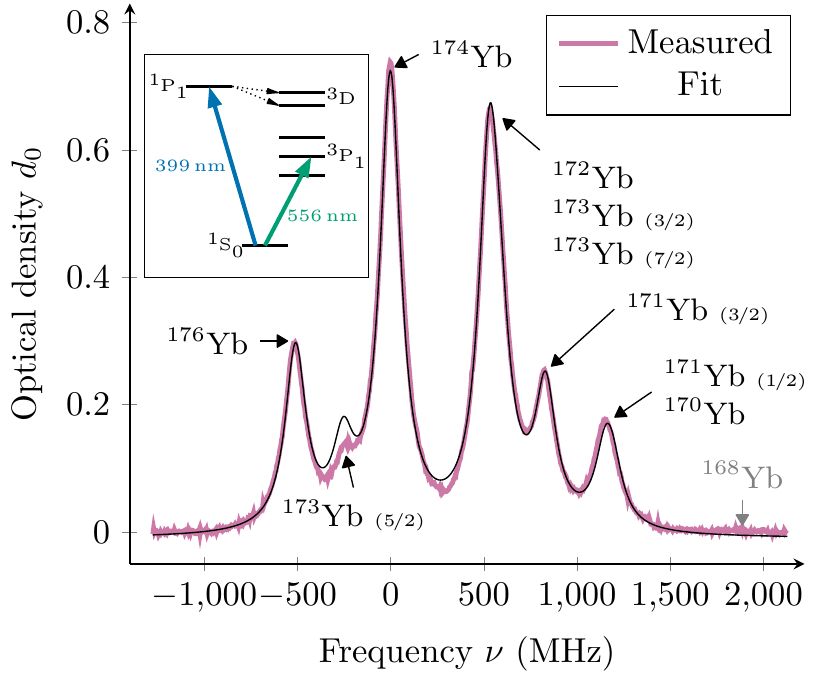}
  \caption{Laser absorption spectrum recorded around the \singletterms resonance at \SI{399}{\nano\meter} in atomic ytterbium. The laser probe beam is perpendicular to the atomic beam (see figure~\ref{fig:oven-setup}), such that the spectral width results from the natural ($2\pi\cdot\Gamma$) and transverse Doppler ($2\pi\cdot\sigma$) widths. We identify six main features, corresponding to nine electronic transitions. For fermions, the number in parentheses indicates the total atomic angular momentum $F$ in the excited state. The boson \Yb{168} is not resolved due to its low natural abundance. \textbf{Inset}: simplified energy level diagram for bosonic atomic ytterbium.}
  \label{fig:oven-spectrum}
\end{figure}

We determine the divergence and flux of the atomic beam shaped by the microchannel nozzle using laser absorption spectroscopy around the \singletterms resonance at \SI{399}{\nano\meter} (figure~\ref{fig:oven-setup}, inset figure~\ref{fig:oven-spectrum}). We use a low saturation ($s = \nicefrac{I}{\Isat} = 0.3$) laser probe of $\nicefrac{1}{e^2}$ diameter $2r_b = \SI{7.5}{\milli\meter}$ directed perpendicular to the atomic beam, around \SI{10}{\centi\meter} after the nozzle. Figure~\ref{fig:oven-spectrum} shows a typical laser absorption spectrum. The six main features correspond to nine resonances in bosonic (\Yb{170}, \Yb{172}, \Yb{174}, \Yb{176}) and fermionic (\Yb{171} ($F'=\nicefrac{1}{2}$, $F'=\nicefrac{3}{2}$), \Yb{173} ($F'=\nicefrac{3}{2}$, $F'=\nicefrac{5}{2}$, $F'=\nicefrac{7}{2}$)) ytterbium \cite{Deilamian1993}. The stable boson \Yb{168} cannot be resolved due its low natural abundance (\SI{0.3}{\percent}) and signal-to-noise ratio limitations in this setup.

To obtain quantitative information about the atomic beam, we adjust the sum of nine Lorentzian profiles of full width at half maximum $\Gamma + 2\tilde{\sigma}$ to the spectrum of figure~\ref{fig:oven-spectrum}. $\Gamma = (2\pi)\cdot\SI{29}{\mega\hertz}$ is the natural width of the \singletterms electronic transition and $\tilde{\sigma}$ quantifies the spectral broadening that results from residual velocity components along the probe laser beam. We found a Lorentzian shape to fit the data better than a Gaussian or Voigt profile. In the adjustment, the spectral positions of the features as well as their relative amplitudes are fixed, respectively set from tabulated values \cite{Das2005}, and calculated from the isotopic natural abundances and standard $L$-$S$ coupling theory \cite{DeglInnocenti2014}. This leaves only five parameters free for the adjustment: the Doppler broadening $\tilde{\sigma}$, a global amplitude quantifying the attenuation of the laser probe by the atomic beam, and three technical parameters that are not physically relevant (global offsets in signal and frequency ranges and frequency ramp calibration). For the spectrum in figure~\ref{fig:oven-spectrum}, we find $\tilde{\sigma} = (2\pi)\cdot\SI{53}{\mega\hertz}$ which corresponds to a transverse velocity $\tilde{v}_t = \SI{21}{\meter\per\second}$ or a half-opening angle $\thetahalftilde = \SI{69}{\milli\radian}$.

We estimate the atomic flux for all isotopes by integrating absorption spectra $d_0(\nu)$ like the one in figure~\ref{fig:oven-spectrum}. Normalizing by the spectrally-integrated scattering cross-section gives the linear density of atoms along the absorption column. Multiplying the column density by the absorption surface $\pi r_b^2$ leads to the instantaneous number of atoms $N_{\mathrm{atoms}}$ in the overlap volume of the laser and atomic beams:

\begin{equation}
  \label{eq:oven-exp-num-scatterers}
  N_{\mathrm{obs}} = \dfrac{\int_{-\infty}^\infty d_0(\nu)\,\mathrm{d}\nu}{\int_{-\infty}^\infty \frac{\sigma_0}{1+4\left[\nicefrac{(2\pi\nu)}{\Gamma}\right]^2}\,\mathrm{d}\nu}\times \pi r_b^2
\end{equation}
where $r_b$ is the probe beam's $\nicefrac{1}{e^2}$ radius and $\sigma_0 = \nicefrac{3\lambda^2}{(2\pi)}\approx \SI{0.08}{\micro\meter\squared}$ is the on-resonance scattering cross section \cite{Metcalf1999}. For a probe beam perpendicular to the atomic beam, the transit time of atoms with longitudinal velocity $\bar{v}$ is on the order of $\nicefrac{2r_b}{\bar{v}} \approx \SI{25}{\micro\second}$. If the probe beam is large enough to intersect the entire atomic beam, the total flux of atoms emerging from the oven is

\begin{equation}
  \label{eq:oven-exp-flux}
  \dot{N} = N_{\mathrm{obs}}\times\frac{\bar{v}}{2 r_b}
\end{equation}
where we estimate $\bar{v}$ by averaging over a Maxwell-Boltzmann distribution at the crucible temperature \Toven.

\begin{figure}
  \centering
  \includegraphics[width=\linewidth]{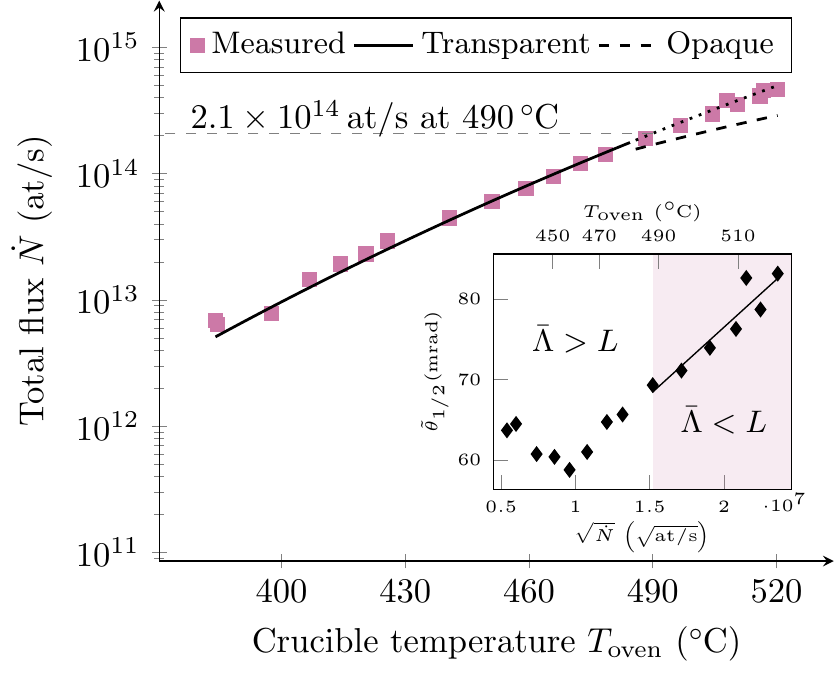}
  \caption{Flux of ytterbium atoms $\dot{N}$ emerging from the microchannel nozzle versus crucible temperature \Toven. Experimental points are derived from absorption spectra applying equation~\eqref{eq:oven-exp-flux}. The solid line corresponds to equation~\eqref{eq:oven-total-flux-trunc-model}, truncated at $\thetamax$ (see figure~\ref{fig:oven-setup}). The dotted line continues that model for temperatures above $\SI{485}{\celsius}$ where the average mean free path $\bar{\Lambda}$ (equation~\ref{eq:oven-avg-mean-free-path}) is smaller than the microchannel length $L$. The dashed curve corresponds to the opaque source model. \textbf{Inset}: onset of the interacting regime. When $\bar{\Lambda}\lesssim L$, the spectral half-width $\thetahalftilde$ scales linearly with the square-root of the measured flux \cite{Giordmaine1960}.}
  \label{fig:oven-flux}
\end{figure}

 Figure~\ref{fig:oven-flux} shows the variation of flux with the crucible temperature \Toven. At our typical operation temperature of \SI{490}{\celsius}, the flux of ytterbium atoms passing the spectroscopy zone exceeds \SI{2e14}{\atom\per\second}.

\subsection{Discussion}

The fit to the spectrum in figure~\ref{fig:oven-spectrum} implies that the atomic beam diverges with a half-angle $\thetahalftilde\approx\SI{70}{\milli\radian}$, much larger than the angle predicted from the collisionless theory for the microchannel nozzle \cite{Ramsey1956,Giordmaine1960} $\thetahalf = \nicefrac{1.68 a}{L} \approx \SI{20}{\milli\radian}$. This suggests a non-negligible influence of interatomic collisions in the nozzle, as evidenced by the mean free path calculation below.

We evaluate the vapor pressure of ytterbium at $\Toven = \SI{490}{\celsius}$ to \SI{2e-2}{\milli\bar} using a Clausius-Clapeyron-type law and tabulated values for the sublimation latent heat \cite{CRCHandbook1997}. Assuming an ideal gas, the corresponding density is $n_0\approx\SI{2.2e20}{\atom\per\cubic\meter}$ and the mean velocity $\bar{v}\approx\SI{300}{\meter\per\second}$. The molecular-flow conductance of the nozzle piece is only \SI{0.02}{\liter\per\second}. This supports five orders of magnitude pressure difference between the entrance and the exit of the microchannels. We assume a constant gradient density profile $n(z) = n_0(1-\nicefrac{z}{L})$ along the microchannels' length $0\leq z\leq L$ \cite{Giordmaine1960}. Let $\bar{\sigma}\approx\SI{400}{\pico\meter}$ the effective atomic diameter of ytterbium. Then, the average mean free path for atoms inside the nozzle capillaries reads:

\begin{equation}
  \label{eq:oven-avg-mean-free-path}
  \bar{\Lambda} = L\int_0^L\frac{\mathrm{d}z}{\Lambda(z)} \ \text{with}\ \Lambda(z) = \left(\pi\sqrt{2} n(z)\bar{\sigma}^2\right)^{-1}.
\end{equation}
We find $\bar{\Lambda}=\SI{11}{\milli\meter}$ at $\Toven = \SI{490}{\celsius}$, which is indeed on the order of the microchannels' length $L = \SI{12}{\milli\meter}$. However, equation~\eqref{eq:oven-avg-mean-free-path} gives $\bar{\Lambda}=\SI{36}{\milli\meter}$ for $\Toven = \SI{450}{\celsius}$ and $\bar{\Lambda} = \SI{5}{\milli\meter}$ at $\Toven = \SI{520}{\celsius}$, indicating a change of regime across the temperature range in figure~\ref{fig:oven-flux}, going from transparent ($\bar{\Lambda}\gg L$) to opaque ($\bar{\Lambda}\ll L$) channels.

Giordmaine and Wang \cite{Giordmaine1960} derive atomic beam angular distributions $J(\theta)$ for both regimes. The total flux follows by integration over the forward half solid angle. However, according to figure~\ref{fig:oven-setup}, the vacuum tubing between the nozzle piece and the spectroscopy region truncates the atomic beam's solid angle at the spectroscopy zone to a polar angle $\thetamax = \SI{6.1}{\degree}$. The prediction for the flux measured by absorption spectroscopy therefore reads:

\begin{equation}
  \label{eq:oven-total-flux-trunc-model}
  \dot{N} = 2\pi\int_0^{\thetamax}J(\theta)\sin{\theta}\,\mathrm{d}\theta
\end{equation}
The solid and dotted lines in figure~\ref{fig:oven-flux} correspond to this model. At the highest temperature probed, the average mean free path is less than half the length of the microchannels. Nevertheless, the collisionless model (dotted line) still qualitatively reproduces the data better than the opaque source model (dashed line). The truncation represents a loss of flux of around an order of magnitude but the geometric acceptance angle of the full Zeeman slower, around \SI{20}{\milli\radian}, is far smaller than \thetamax and therefore constitutes the major constraint on useable flux.

Despite the qualitative agreement for the total flux, the theoretical prediction doesn't match the observed half-width $\thetahalftilde$ quantitatively. The half angle divergence in the collisionless theory is around \SI{20}{\milli\radian} and above \SI{100}{\milli\radian} in the opaque theory, compared to measured values between \SI{60}{\milli\radian} and \SI{80}{\milli\radian}. This mismatch is likely due to the mixed regime in which we operate the nozzle. The inset in figure~\ref{fig:oven-flux} shows some evidence for the opaque regime at higher temperatures where the beam half-width scales with the square root of the total flux, while it is predicted to be independent of flux in the transparent regime. Nevertheless, the truncation of the atomic beam by the vacuum chamber prevents a faithful observation of the angular distribution, although not limited by a thin aperture as in other work \cite{Schioppo2012}. A study of the onset of the opaque regime is of high interest for future applications of high flux ovens, for example in view of gravitational wave detection with atomic test masses \cite{Loriani2019}. Cascaded designs \cite{Li2019} potentially enable higher collimation and better recycling of atoms with highly diverging trajectories. Also, in highly opaque configurations, even moderate optical recollimation can be effective at the exit of the oven \cite{Yang2015}.

%% file: sec-zs.tex
At $\Toven = \SI{490}{\celsius}$, the oven and its microchannel nozzle produce a beam of around \SI{2.1e14}{\atoms\per\second} travelling at a mean forward velocity $\bar{v}\approx\SI{300}{\meter\per\second}$. We employ Zeeman slowing to reduce the forward velocity to fit the few tens of \si{\meter\per\second} capture velocity of our 2D-MOT operating around the \singletterms transition. We focus on the most abundant isotope, \Yb{174}.

We set up rare-earth magnets in a Halbach configuration to generate the magnetic field profile required for Zeeman deceleration. This approach is straightforward to assemble and requires neither high-power electrical supplies nor water cooling, thus increasing reliability \cite{Hill2016,Ovchinnikov2007}. Also, the Halbach configuration provides efficient suppression of stray fields.

In this section, we briefly review elements of Zeeman slowing theory and motivate our design parameters (section~\ref{sec:zs-theory}). We then describe our permanent magnet layout in depth and characterize its magnetic properties (section~\ref{sec:zs-halbach}). Finally, we evaluate the performance of the slower on the atomic beam (section~\ref{sec:zs-performance}).

\subsection{Zeeman slower design}
\label{sec:zs-theory}

Zeeman slowing \cite{Phillips1982} exploits the fact that atoms can be decelerated when scattering photons from a counter-propagating laser beam. However, the change in velocity leads to a varying Doppler shift that needs to be accounted for to maintain the resonance condition for the scattering process. This is achieved using the shift of electronic energy levels arising from the coupling of an external magnetic field to the electronic angular momentum, the so-called ``Zeeman effect''.

The theory of Zeeman slowing is extensively covered in the literature \cite{Metcalf1999}. We review the essential aspects to fix the notation. We consider atomic two-level systems with an zero-field energy splitting $\hbar\omega_0$ travelling at velocity $v(z)$ along an axis labelled by coordinate $z$. The differential linear Zeeman shift between the energy levels when the atom is at position $z$ is $\tilde{\mu} B(z)$ where $|\tilde{\mu}|$ equals \num{1.035} Bohr magnetons for bosonic ytterbium atoms on the $^1S_0\to{}^1P_1$ transition \cite{Dareau2015}. In frequency units, this amounts to \SI{\pm 14}{\mega\hertz\per\milli\tesla}. A laser beam with frequency $\nicefrac{\omega}{(2\pi)}$ is directed against the atomic beam. At position $z$, the detuning $\delta$ to the atomic resonance reads: \cite{Hopkins2016}

\begin{equation}
  \label{eq:zs-detuning}
  \delta(z) = \Delta + kv(z) - \frac{\tilde{\mu}B(z)}{\hbar}
\end{equation}
where $\Delta = \omega - \omega_0$ is the laser detuning to the resonance for atoms at rest, $k = \nicefrac{2\pi}{\lambda}$, the light's wavenumber, and $\hbar$ the reduced Planck constant.

Setting $\delta(z) = 0$ for the whole length of the Zeeman slower, the equation of motion of atoms in this region follows from standard scattering theory:

\begin{equation}
  \label{eq:zs-eom}
  \frac{\mathrm{d}v}{\mathrm{d}t} = -\underbrace{\frac{s}{1+s}}_{\eta}\times\underbrace{\frac{\hbar k\Gamma}{2m}}_{\amax}
\end{equation}
with $s$ the saturation parameter, $m$ the atomic mass and $\Gamma$ the natural width of the transition. The parameter $\eta < 1$ quantifies the fraction of the scattering rate-limited maximum deceleration $-\amax$ required for the atoms to stay on resonance and continue on the slowing trajectory. A solution of equation~\eqref{eq:zs-eom} reads

\begin{equation}
  \label{eq:zs-eom-sol-v}
  v(z) = v_c\cdot\sqrt{1-\dfrac{z}{\Lstop}}
\end{equation}
where $v_c = v(z=0)$ is the capture velocity (the largest velocity class addressed by the slowing process), and $\Lstop = \nicefrac{v_c^2}{2\eta\amax}$ is the distance required to bring the atoms to rest with the constant deceleration $-\eta\amax$. It is however necessary for the atoms to be extracted from the slower that they have a small exit velocity $v_e$. The slowing length required to reach $v_e$ is $L = \Lstop - \nicefrac{v_e^2}{2\eta\amax} < \Lstop$.

Replacing equation~\eqref{eq:zs-eom-sol-v} in equation~\eqref{eq:zs-detuning} for $\delta(z) = 0$ and solving for the magnetic field profile $B(z)$, one finds:

\begin{equation}
  \label{eq:zs-ideal-field}
  B(z) =
  \begin{cases}
    B_0 + B_L\cdot\sqrt{1-\dfrac{z}{\Lstop}} & 0 < z < L \\
    0 & \text{otherwise}
  \end{cases}
\end{equation}
The design degrees of freedom are the field offset $B_0 = \nicefrac{\hbar\Delta}{\tilde{\mu}}$, and the amplitude $B_L = \nicefrac{\hbar k v_c}{\tilde{\mu}}$. The field profile is truncated at $z=L<\Lstop$ to be able to extract the slow atoms. The field therefore varies from $B_0 + B_L$ at $z=0$ to $B_0 + \nicefrac{\hbar k v_e}{\tilde{\mu}}$ at $z=L$. The exit velocity $v_e$ can be experimentally tuned either by adjusting the magnitude of the field near the end of the slower, or by varying the detuning $\Delta$. We design the slower for $\Lstop$, not forgetting that slight adjustments in detuning and field magnitude near the end will be necessary.

Table~\ref{tab:zs-parameters} summarizes the design parameters for our Zeeman slower. We set the capture velocity $v_c = \SI{390}{\meter\per\second}$ that corresponds to around \SI{75}{\percent} of the atoms in the beam with a forward velocity below the slowing threshold. The minimal slower length is \SI{14}{\centi\meter} ($\eta = 1$). We choose $\Lstop=\SI{30}{\centi\meter}$, i.e. a deceleration margin parameter $\eta\approx\num{0.5}$. The detuning $\Delta$ determines the offset $B_0$ and hence the magnitude of the maximum field to generate. Minimizing this quantity would lead to the choice $B_0 = -\nicefrac{B_L}{2}$, that is $\Delta = -2\pi\cdot\SI{490}{\mega\hertz}$. However, for geometrical reasons, the Zeeman slowing laser beam passes through, or very close to, the center of the subsequent 2D-MOT which needs not to be unbalanced or lifetime limited through excessive losses to the $\term{3}{D}{}$ states (see inset figure~\ref{fig:oven-spectrum}). We choose $\Delta = -2\pi\cdot\SI{700}{\mega\hertz}$, which corresponds to a loss time constant above \SI{200}{\second} \cite{Freytag2015}. Also, the influence of the Zeeman slowing beam on the scattering rate is more than \num{1000} times smaller than that of the 2D-MOT beams. This results in a maximum field amplitude of \SI{40}{\milli\tesla}. The magnetic field profile corresponding to equation~\ref{eq:zs-ideal-field} with the parameters in table~\ref{tab:zs-parameters} is shown as the solid green line in the middle panel in figure~\ref{fig:zeeman-slower}.

\begin{table}
	\tablestyle
	\renewcommand{\arraystretch}{2.2}
	\centering
	\begin{tabularx}{0.95\linewidth}{
			>{\columncolor{white}[\tabcolsep][\tabcolsep]}L{0.3\linewidth}%
			>{\columncolor{white}[\tabcolsep][\tabcolsep]}L{0.3\linewidth}%
                        >{\columncolor{white}[\tabcolsep][\tabcolsep]}L{0.3\linewidth}}%
		\theadstart
		\thead Parameter &
		\thead Symbol &
                \thead Value
		\tabularnewline
		\theadend
		\tbody
		Angular momentum change &  $\tilde{\mu}$ & $\num{-1.035}\cdot\mu_B$ \\
		Maximum deceleration & $-\amax = -\dfrac{\hbar k\Gamma}{2m}$ & \SI{-5.3e5}{\meter\per\second\squared} \\
		Detuning & $\Delta$ & $-2\pi\cdot\SI{700}{\mega\hertz}$ \\
		Capture velocity & $v_c$ & \SI{390}{\meter\per\second} \\
		Length & $\Lstop$ & \SI{30}{\centi\meter} \\
		\hline
		Deceleration margin & $\eta = \dfrac{v_c^2}{2\Lstop\amax}$ & \num{0.48} \\
		Saturation parameter & $s=\dfrac{\eta}{1-\eta}$ & \num{0.93} \\
		Field amplitude & $B_L = \dfrac{\hbar k v_c}{\tilde{\mu}}$ & \SI{-67}{\milli\tesla} \\
		Field offset & $B_0 = \dfrac{\hbar\Delta}{\tilde{\mu}}$ & \SI{48}{\milli\tesla} \\
		\tend
	\end{tabularx}
	\caption{Zeeman slower design parameters}
	\label{tab:zs-parameters}
\end{table}

\subsection{Zeeman slowing field from permanent magnets in Halbach array configuration}
\label{sec:zs-halbach}

\begin{figure*}
  \centering
  \includegraphics[width=\linewidth]{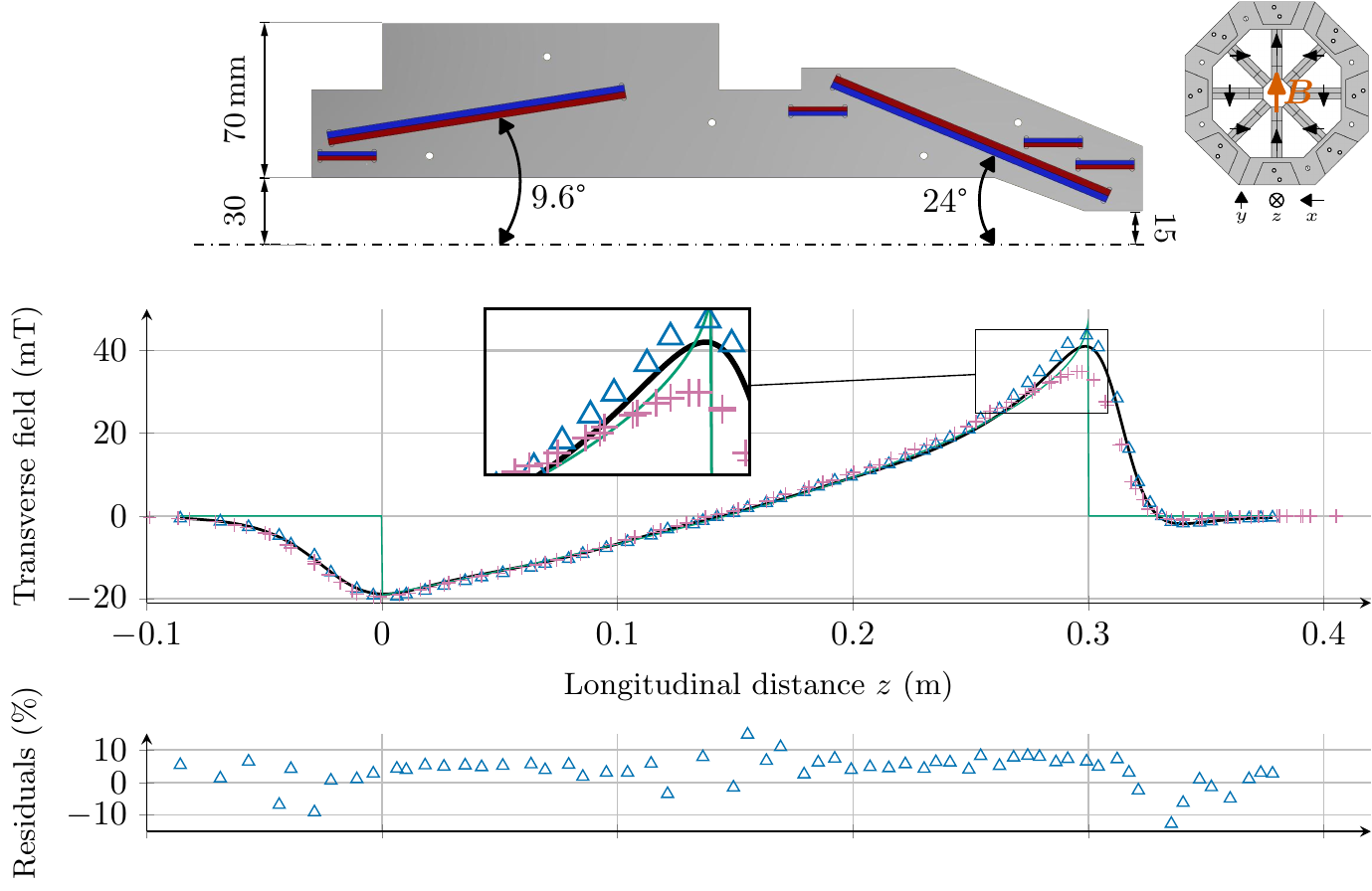}
  \caption{Construction and magnetic field of the permanent magnets-based Zeeman slower. \textbf{Top panel}: mechanical arrangement of the two main magnets and four trimmer magnets in one of the eight blades forming the magnet assembly. The inset shows the magnetization direction for the entrance trimmer magnet on all eight blades, effectively forming an eight-fold discrete Halbach configuration. \textbf{Middle panel}: ideal (\showline{cud-bluish-green}{0.7pt}), calculated (\showline{black}{1pt}), and measured (\showmark{triangle}{cud-blue}, \showmark{+}{cud-reddish-purple}) strong transverse component of the magnetic field along the slower's magnet. \showmark{+}{cud-reddish-purple} represents the externally trimmed configuration used for loading the 2D-MOT of section~\ref{sec:2d-mot}. \textbf{Lower panel}: residuals of the measurements (\showmark{triangle}{cud-blue}) to the corresponding calculated configuration (\showline{black}{1pt}). Most points lie within \SI{10}{\percent} of the numerical model.}
  \label{fig:zeeman-slower}
\end{figure*}

We use permanent magnets in a Halbach configuration to produce the magnetic field described by equation~\eqref{eq:zs-ideal-field}. Halbach arrays are particular magnetic configurations that produce magnetic multipole fields in a well constrained domain in space, while suppressing the field outside this region \cite{Halbach1980}. We consider a magnetic cylinder with large length-to-diameter ratio and prescribe that the material's magnetization $M$ rotates as follows in the plane ($\unitv{\rho},\,\unitv{\phi}$) transverse to the cylinder's axis:

\begin{equation}
  \label{eq:halbach-cylinder-dipole}
  \frac{\mathbf{M}}{|\mathbf{M}|} \equiv\unitv{M} = \cos\left(2\phi\right)\unitv{\rho} + \sin\left(2\phi\right)\unitv{\phi}.
\end{equation}
The corresponding magnetic field is transverse and homogeneous along $+\unitv{\rho}$ and zero outside the cylinder. If the magnetic object is not a cylinder but a cone, the magnitude of the transverse field decreases with increasing cone radius, however also introducing a longitudinal field component except on the cone's axis. We assert numerically, as already shown in previous work \cite{BenAli2017,Cheiney2011}, that a Halbach cone produces a percent-level approximation of the ideal Zeeman slowing field from equation~\eqref{eq:zs-ideal-field}.

Since a continuous magnetic material with the magnetization rotation described by equation~\eqref{eq:halbach-cylinder-dipole} is not practical, and similarly to the work in references \cite{BenAli2017,Cheiney2011}, we discretize the array using an eight-fold symmetry. However, contrary to previous work, our design exhibits a zero crossing and a  higher overall average gradient (\SI{200}{\milli\tesla\per\meter} versus \SI{120}{\milli\tesla\per\meter} for Ref.~\cite{BenAli2017}, and \SI{33}{\milli\tesla\per\meter} for Ref.~\cite{Cheiney2011}).

The top panel in figure~\ref{fig:zeeman-slower} shows the construction of our magnet. The main field is generated by two sets of eight \ch{NdFeB} rectangular cuboids \footnote{HKCM Engineering, part no. Q128x06x06Zn-30SH} with length \SI{128}{\milli\meter}, $\SI{6}{\milli\meter}\times\SI{6}{\milli\meter}$ square cross-section, and nominal remanence $B_r= \SI{1.08}{\tesla}$, magnetized along one of the transverse directions. These long cuboids are arranged in two Halbach cones with slopes $+\SI{9.6}{\degree}$ and $-\SI{24}{\degree}$, generating the main part of the downwards and upwards pointing transverse field (figure~\ref{fig:zeeman-slower}, middle panel). We additionaly use $\num{8}\times\num{4}=\num{32}$ smaller rectangular cuboids ($\SI{25}{\milli\meter}\times\SI{4}{\milli\meter}\times\SI{4}{\milli\meter}$, $B_r = \SI{1.17}{\tesla}$ \footnote{HKCM Engineering, part no. Q25x04x04Zn-35H}) as adjustment variables to better fit the starting field, the zero-field transition regime in the middle of the slower, and help decreasing the field faster at the exit of the slower. We numerically optimize the positions and orientations of all the permanent magnets, while enforcing the Halbach symmetry. We use the analytic expression derived in Ref.~\cite{Cheiney2011} to calculate the field from each magnet. The optimized configuration leads to the field profile reprensented by the solid black curve in the middle panel in figure~\ref{fig:zeeman-slower}.

The full magnet consists of eight sets of six magnets clamped between aluminium blades. The positions of the magnets on the plates is engraved (e.g. \SI{3}{\milli\meter} deep per plate for the long magnets) during the manufacturing process. The assembly of the magnet takes less than two hours, dominated by the identification of the magnetization direction for the \num{48} permanent magnets. We kept the manufacturing tolerances below \SI{0.5}{\milli\meter} to constrain the assembly sufficiently while taking size variations of the individual magnets into account. The plates are pressed against each other using five M3 bolts and mounted in an octagonal holder (figure~\ref{fig:zeeman-slower}, inset) to obtain the desired configuration. The total weight of this assembly is \SI{10}{\kilo\gram}. We measure the magnetic field along the magnet's axis using a 3-axes teslameter \footnote{F.W. Bell, Model 7030}. As shown in figure~\ref{fig:zeeman-slower} (middle panel: blue triangles), we reproduce the calculated field (solid black line) to around \SI{10}{\percent} (lower panel). In the transverse direction, the field decays to background values within \SI{10}{\centi\meter} outside the magnet assembly.

\subsection{Slowing performance}
\label{sec:zs-performance}

Our design achieves slowing by scattering photons on the $\term{1}{S}{0}\to\term{1}{P}{1}\,(m_J=-1)$ transition, which requires $\sigma^-$ polarized light to account for the change in electronic angular momentum. However, since the generated field is transverse to the atomic propagation direction, the configuration maximizing the $\sigma^-$ polarization content corresponds to light polarized linearly along \unitv{x}, perpendicular to the magnetic field direction \unitv{y} (see inset of figure~\ref{fig:zeeman-slower}). This however implies that only half of the optical power has $\sigma^-$ polarization and contributes to the dominant slowing effect. The $\sigma^+$ polarization component is resonant with the atoms at some positions in the slower but this doesn't affect overall performance. An influence via (coherent) population dynamics is ruled out by the non-magnetic \term{1}{S}{0} ground state. The slowing beam has a $\nicefrac{1}{e^2}$ diameter of \SI{4}{\milli\meter} and total power of \SI{80}{\milli\watt} resulting in a peak intensity of \SI{3.2e2}{\milli\watt\per\square\centi\meter} ($\approx 2.7\times\Isat$ with $\sigma^-$ polarization).

Coupling the slowing light beam inside the vacuum chamber requires special care. Since the slowing beam is anti-parallel to the atomic beam, an optical surface must be facing the atomic beam directly. A common solution to reduce metallic deposition is to use an uncoated, z-cut sapphire viewport heated to several hundreds of degrees celsius \cite{Stellmer2013}. We found this inconvenient, namely because of the high maintenance cost and added heat source near the main experimental chamber. An alternative solution is an in-vacuum protected aluminum mirror, as demonstrated in reference~\cite{Huckans2018} (see inset in figure~\ref{fig:setup}). While the chemical mechanism for maintaining high reflectivity under metallic deposition is not well-known, operation of our atomic beam over several months evaporated more than \SI{100}{\milli\gram} of ytterbium from the oven with neither performance loss nor visible degradation of the mirror surface.

\begin{figure}
  \centering
  \includegraphics[width=\linewidth]{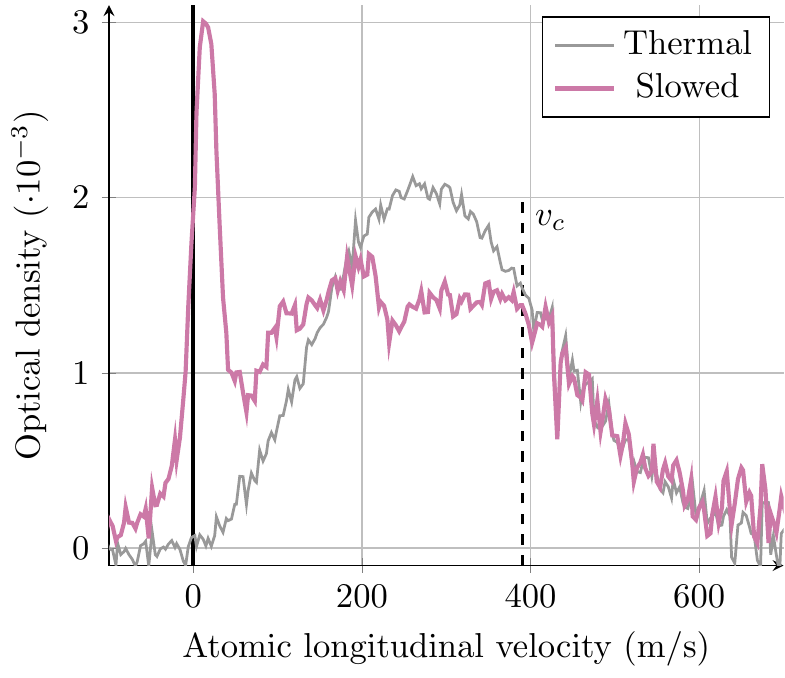}
  \caption{Absorption spectroscopy of the atomic beam  under a \SI{30}{\degree} angle at the exit of the Zeeman slower. The resulting spectrum shows the longitudinal velocity distribution and the effect of the slowing laser beam. Atoms exiting the oven with a velocity below $v_c \approx \SI{390}{\meter\per\second}$ are slowed down to below \SI{50}{\meter\per\second}. The frequency axis is calibrated using a simultaneous spectroscopy measurement under an axis perpendicular to the atomic beam.}
  \label{fig:slowing-curve}
\end{figure}

We characterize the slowed atomic beam using absorption spectroscopy under a \SI{30}{\degree} angle at the exit of the slower. Figure~\ref{fig:slowing-curve} shows the longitudinal velocity profile for \Yb{174} atoms. The capture velocity around \SI{390}{\meter\per\second} is clearly visible, as well as the collection of low velocity ($<\SI{50}{\meter\per\second}$) atoms. We maximize the flux of slow atoms by adjusting the laser detuning to the \singletterms transition and tuning the magnetic field with two additional trimming bar magnets placed symmetrically at the exit of the magnet assembly. The resulting field is plotted using pink crosses in the middle panel in figure~\ref{fig:zeeman-slower}. The main effect of these extra trimmers is to avoid significantly overshooting the ideal magnetic field profile, thus keeping the local deceleration parameter $\eta$ close to its design value and therefore avoid loosing atoms from the slowing profile due to missing light intensity.

For the data presented in figure~\ref{fig:slowing-curve}, the detuning is $\Delta = -(2\pi)\cdot\SI{580}{\mega\hertz}$, $4\Gamma$ away from the design value $-(2\pi)\cdot\SI{700}{\mega\hertz}$. This is caused by the decrease in field magnitude due to the external trimmers, and the need for a finite exit velocity. The most probable exit velocity in figure~\ref{fig:slowing-curve} is $v_e = \SI{15}{\meter\per\second}$. We measured a trimmed field maximum of \SI{35}{\milli\tesla}, which matches reasonably the prediction from equation~\eqref{eq:zs-ideal-field} $\nicefrac{\hbar\Delta}{\tilde{\mu}} + \nicefrac{\hbar k v_e}{\tilde{\mu}} = \SI{37}{\milli\tesla}$.

Ideally, all atoms below the capture velocity should be slowed down. This is evidently not the case in figure~\ref{fig:slowing-curve}, mainly due to the divergence of the atomic beam inside the slower. Atoms travelling too far off-axis are not interacting with the slowing beam and therefore not decelerated further. This effect is amplified with decreasing velocity since the longitudinal velocity becomes comparable to the transverse one. Increasing the slowing beam diameter or decreasing the atomic beam divergence with radiation pressure prior to the Zeeman slower \cite{Yang2015} can effectively mitigate this issue, at the cost of increased optical power requirements.

%% file: sec-2d-mot.tex
The divergence of the atomic beam during Zeeman slowing is a major limitation for the flux of slow atoms capturable by a 3D-MOT. Moreover, constraints on the target apparatus for this atomic source dictate a large size of the main experimental vacuum chamber, leading to a minimum distance of more than \SI{30}{\centi\meter} between the exit of the slower and the center of the 3D trap. We therefore have to keep the atoms' longitudinal velocity large compared to their transverse velocity in order to enable efficient loading of the 3D-MOT.

In order to mitigate the beam divergence issue, we implement a 2D-MOT between the Zeeman slower and the 3D-MOT. This presents the following advantages. First, it efficiently recollimates the atomic beam, preventing significant losses even when passing through a \SI{4}{\milli\meter} diameter differential pumping aperture between the 2D-MOT and 3D-MOT chambers. Second, since we load it under a \SI{30}{\degree} angle, it acts as a dump for fast, non slowed atoms, preventing them from entering the main chamber. Finally, the Zeeman slowing beam can be coupled through the 2D-MOT chamber, thus reducing the complexity around the 3D-MOT chamber and avoiding perturbing the 3D-MOT with the slowing light (see figure~\ref{fig:setup}).

\subsection{Design}

\begin{figure}
  \centering
  \includegraphics[width=\linewidth]{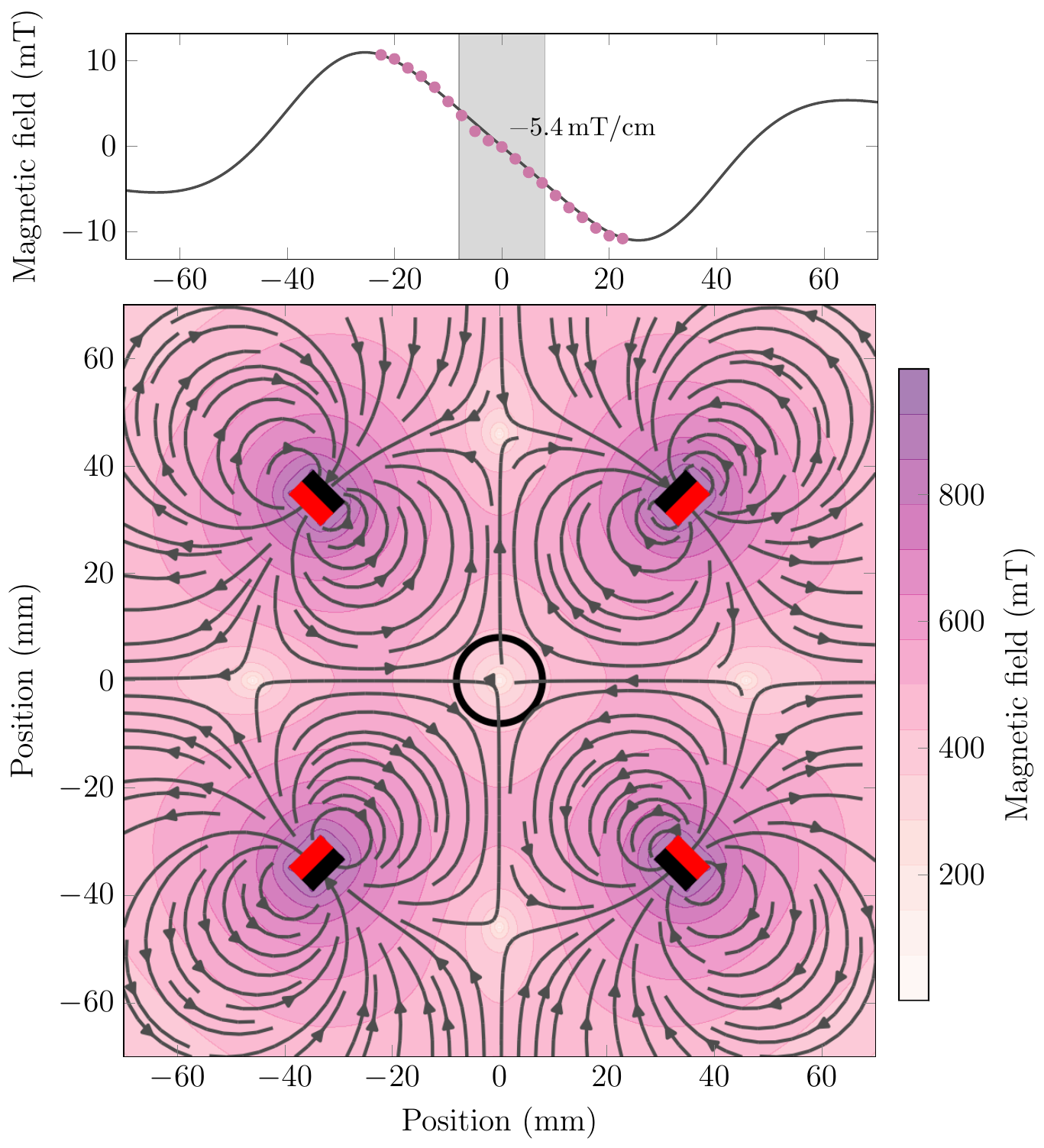}
  \caption{Permanent magnet configuration and transverse field profile for the 2D-MOT generated with permanent magnets. Top:~simulation \showline{black}{0.7pt} and measurement \showmark{*}{purple!60!gray} showing a gradient of (-)\SI{5.4}{\milli\tesla\per\cm}. The shaded area indicates the interior of the vacuum chamber. Bottom: simulation of the transverse cross-section of the field profile. The red/black rectangles indicate the positions and orientations of the four permanent bar magnets with red(black) indicating their north(south) pole. The black circle represents the inner wall of the vacuum chamber.}
  \label{fig:2d-mot}
\end{figure}

We follow the same design guidelines for the magnetic configuration of the 2D-MOT as for the Zeeman slower and use four permanent magnet bars ($\SI{80x8x6}{\milli\meter}$, $B_r=\SI{1.17}{\tesla}$ \footnote{HKCM Engineering Q80x08x06Zn-35H})) in a four-pole Halbach configuration \cite{Tiecke2009} to produce around $\SI{5.5}{\milli\tesla\per\centi\meter}$ gradients. Figure~\ref{fig:2d-mot} shows the magnetic configuration, as well as simulated and measured magnetic field profiles. Our design includes electromagnetic coils for fine tuning the position of the quadrupole zero but we do not use them for the following results. Therefore, like for the Zeeman slower, the magnetic field for the 2D-MOT is generated permanently and passively, increasing the robustness and reliability of the system.

We generate the optical radiation pressure using two retro-reflected $\SI{1}{\centi\meter}\times\SI{4}{\centi\meter}$ laser beams with \SI{40}{\milli\watt} each ($\approx\num{0.4}\times\Isat$) tuned \SI{16}{\mega\hertz} ($0.5\Gamma$) to the red of the \singletterms resonance. Beam shaping is achieved using pairs of cylindrical lenses.

\subsection{Characterization}

We characterize the atomic beam produced by the 2D-MOT with retro-reflected spectroscopy in the main chamber. Since the 2D-MOT mostly affects velocity components perpendicular to its axis, the longitudinal velocity of atoms exiting the 2D-MOT corresponds to the projection of the slower's exit velocity on the 2D-MOT axis. We find a flux optimum for a 2D-MOT exit velocity $\bar{v}_l = \SI{20}{\meter\per\second}$, characterizing the trade-off between low output velocity and excessive losses due to beam divergence in the Zeeman slower.

Since the transverse velocity is not resolved in our retro-reflected spectroscopy setup, we deduce it from the size of the atomic beam in the center of the main chamber. Varying the probe beam diameter, we estimate an atomic beam diameter of \SI{1}{\centi\meter} at a distance of \SI{23}{\centi\meter} from the differential pumping pinhole. This corresponds to a maximum transverse velocity $\left(\nicefrac{\SI{1}{\cm}}{\SI{23}{\cm}}\right)\cdot\bar{v}_l < \SI{1}{\meter\per\second}$ or five times the Doppler limit (\SI{0.18}{\meter\per\second}) and represents a factor \num{20} improvement over the atomic beam divergence at the exit of the oven.

The trade-off between divergence and low exit velocity is also relevant at the exit of the 2D-MOT. Indeed, the longitudinal velocity must be small enough to enable capture by the 3D-MOT, while maintaining decent beam collimation despite the \SI{23}{\centi\meter} travel distance between the differential pumping pinhole and the center of the 3D-MOT. Since the transverse velocity is on the order of \SI{1}{\meter\per\second}, matching the few meters per second capture velocity of a typical MOT operating on the intercombination transition \tripletterms ($\Gamma = 2\pi\cdot\SI{180}{\kilo\hertz}$) would result in an atomic beam diameter of around ten centimeters, which is impractical. With \SI{20}{\meter\per\second} however, the atomic beam size is restricted to single digit centimeters while still capturable by a MOT operated on the \singletterms ($\Gamma = 2\pi\cdot\SI{29}{\mega\hertz}$) transition.

%% file: sec-3d-mot.tex
\begin{figure}
  \centering
  \includegraphics[width=\linewidth]{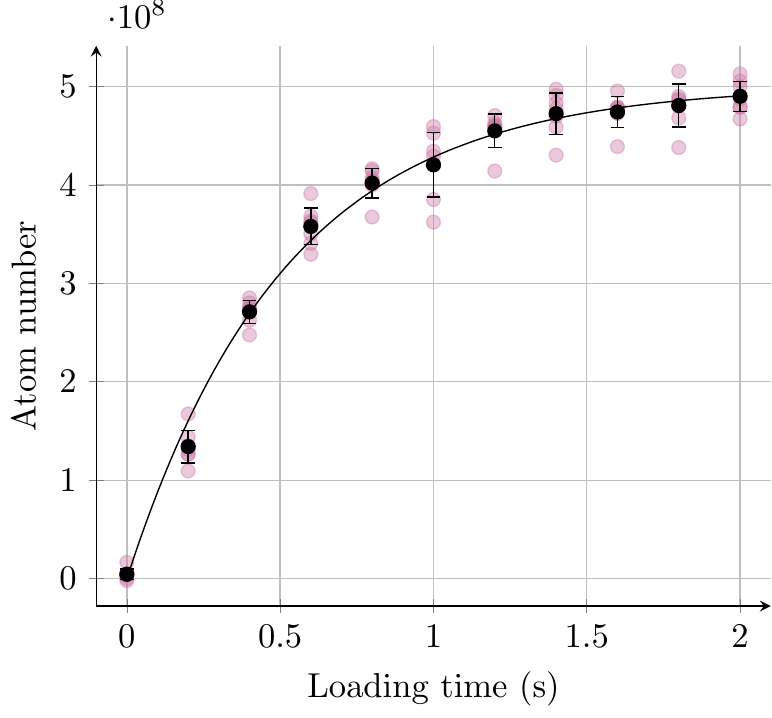}
  \caption{Loading curve of the 3D MOT. The solid line is a fit of eq.~\ref{eq:mot-loading}, with best fit parameters $\gamma = \SI{1e9}{\atom\per\second}$ and $R = \SI{2}{\per\second}$. The errorbars indicate the standard deviation of the measured datapoints.}
  \label{fig:mot-loading}
\end{figure}

We characterize the overall performance of the system consisting of the oven, the Zeeman slower, and the 2D-MOT by capturing the ytterbium atoms in a large volume 3D-MOT and measuring the trap's loading rate. This provides a realistic estimate of the useable flux for further cooling steps but also removes the need for geometrical assumptions associated with spectroscopic flux measurements.

\subsection{3D-MOT}

We operate the 3D MOT on the \singletterms transition. The laser beams have a $\nicefrac{1}{e^2}$ diameter around \SI{2}{\cm}. The magnetic field gradient is produced by a pair of coils in anti-Helmholtz configuration.

We determine the number of trapped atoms by absorption imaging and collection of the fluorescence from trapped atoms. While both methods are in qualitative agreement, we exclusively use absorption imaging for the quantitative results below, as it provides the more conservative values and requires less assumptions.

We optimize the trap for loading rate and maximum atom number. We find the optimal loading performance for a detuning $\Delta = \SI{-32}{\MHz}$ ($-1.1\Gamma$) from the \singletterms resonance, a magnetic field gradient $\delta B = \SI{210}{\milli\tesla\per\meter}$, and $\Pmot = \SI{30}{\milli\watt}$ per beam ($0.08\times\Isat$ per beam). Figure~\ref{fig:mot-loading} shows an average loading curve. Our trap saturates at a steady-state atom number of \SI{5e8}{\atoms} within about \SI{2}{\second}. We extract the loading rate using a one-body loss rate model:

\begin{equation}
  \label{eq:mot-loading}
  N(t) = \frac{\gamma}{R} \left( 1-\exp(-Rt) \right),
\end{equation}
where $\gamma$ is the loading rate. The one-body loss rate $R$ contains losses due to background collisions but is dominated by radiative loss via the \term{3}{D}{1,2} to the \term{3}{P}{0,2} states (inset figure~\ref{fig:oven-spectrum}), making it highly dependent on beam intensity and detuning~\cite{Cho2012}. Despite the relatively high density ($\approx\SI{1e9}{\atom\per\cubic\cm}$) no significant two-body losses were observed compared to the strong one-body loss rate and are therefore neglected.

Adjusting the model of equation~\eqref{eq:mot-loading} to the data in figure~\ref{fig:mot-loading}, we find $\gamma = \SI{1e9}{\atom\per\second}$ and $R = \SI{2}{\per\second}$. At higher powers the maximum loading rate remains the same but relaxes the requirements on detuning and gradient, allowing the same loading rate in a larger part of the parameter space. This behaviour suggests that in these configurations the loading rate is limited only by the flux into the main chamber and would allow for larger steady-state numbers at larger detunings and gradients as long as sufficient laser power can be provided.

\subsection{System performance evaluation}

Starting from a flux of \SI{2e14}{\atom\per\second} at the exit of the oven, our system loads \num{5e8} \Yb{174} atoms in \SI{2}{\second} (\SI{1e9}{\atom\per\s} initial loading rate) in a 3D-MOT on the \singletterms transition. Accounting for the natural abundance of \Yb{174} (\SI{32}{\percent}), this represents a loss of five orders of magnitude in atom numbers during the slowing, redirection, and recollimation processes.

Table~\ref{tab:flux} shows a summary of diagnostic flux measurements performed along the beamline. The major loss contributor is the Zeeman slower. The loss of total flux, cumulating all velocity classes at the exit of the Zeeman slower's vacuum pipe, is accounted for by the slower's length (\SI{60}{\centi\meter} including all connection pieces and isolation valves) and the divergence of the atomic beam exiting the oven (\SI{70}{\milli\radian}, see figure~\ref{fig:oven-flux}). This gives a maximum throughput of \SI{1.3}{\percent}, i.e. a maximum flux of \SI{8e11}{\atom\per\second}, in good agreement with the measured value of \SI{5e11}{\atom\per\second}. The loss of flux during the slowing process is split between the slower's velocity acceptance (\SI{25}{\percent} of the atoms are above the capture velocity $v_c = \SI{390}{\meter\per\second}$) and atomic beam divergence inside the slower (see discussion in section~\ref{sec:zeeman-slower}), with the latter dominating largely.

Reducing the divergence of the atomic beam at the exit of the oven is therefore a good way of improving the slowing efficiency and preserving flux along the beamline. As an example, reducing the divergence half-angle by a factor \num{3} increases the geometric throughput by almost a factor \num{8} and we expect this to also improve on the slowing losses. Since the oven is operated near the opaque channel regime to achieve high flux, the collimation needs to be performed optically. With our apparatus, the above recollimation would require around \SI{150}{\milli\watt} of laser power which are currently not at our disposal. Improvements by simple optical collimations after the oven have been observed by experiments, for example in reference~\cite{Yang2015}. Nevertheless, when the oven nozzle is in the opaque channel regime, the divergence increases with flux, correspondingly constraining the laser system or the geometry of the optical collimation chamber. Alternatively, the slower's length could be reduced, limiting divergence effects but decreasing it's safety margin $\eta$, or the slowing beam's diameter can be increased, at the expense of a considerable rise in laser power consumption.

As discussed in section~\ref{sec:2d-mot}, the current apparatus is unsuitable for direct loading of a MOT operating on the \tripletterms transition~\cite{Guttridge2016}. This is mostly due to the size of the 3D trap chamber, constrained by other design requirements for this apparatus. In order to keep a reasonably sized MOT, the transverse velocity around \SI{1}{\meter\per\second} of the atoms exiting the 2D-MOT require the forward velocity to stay above \SI{20}{\meter\per\second} while not significantly sacrificing flux. In order to benefit from narrow line cooling, further optimization of the 2D-MOT stage up to the Doppler limit is a possibility. However, it is more versatile to add an extra slowing stage in the main chamber. The most straightforward would be sequential loading via a singlet MOT but other schemes such as two-stage cooling~\cite{Lunden2020} or a core-shell MOT~\cite{Lee2015} might also be considered.

\begin{table}
	\tablestyle
	\renewcommand{\arraystretch}{1.5}
	\begin{tabularx}{0.95\linewidth}{
			>{\columncolor{white}[\tabcolsep][\tabcolsep]}X%
			>{\columncolor{white}[\tabcolsep][\tabcolsep]}Y}
		\theadstart
		\thead Position along beamline & \thead Flux in \si{\atoms\per\second} \tabularnewline
		\theadend
		\tbody
		Oven & \num{6e13}{} \\
		Zeeman slower total & \num{5e11}{} \\
		Zeeman slower ($< \SI{50}{\m\per\s}$) & \num{6e9} \\
		Captured by 3D-MOT & \num{1e9} \\
		\tend
	\end{tabularx}
        \caption{Flux of \Yb{174} atoms at different positions along the beamline. The oven flux is given in figure~\ref{fig:oven-flux} weighted by the natural abundance of \Yb{174}. For the Zeeman slower, we distinguish between the total flux of atoms, cumulating all velocity classes, and the slowed flux, accounting only atoms with less than \SI{50}{\meter\per\second} forward velocity. Finally, the captured flux is the loading rate into the 3D-MOT according to equation~\eqref{eq:mot-loading}.}
        \label{tab:flux}
\end{table}

%% file: sec-conclusion.tex
We have built and characterized a source of laser-cooled ytterbium atoms delivering \SI{1e9}{\atoms\per\second} in a 3D magneto-optical trap. All necessary magnetic fields on the atomic beamline are produced by permanent magnets in Halbach configurations. This provides easily reproducible designs with low stray magnetic fields, as well as robustness since no maintenance on e.g. water cooling circuits is required. Also, the resulting assembly is lightweight and can be disassembled for investigation, vacuum bakeout, or transportation. Apart from the laser systems, the only electrical power consumption stems from the oven ($< \SI{50}{\watt}$) and could be further reduced using in-vacuum heating \cite{Schioppo2012}. In particular, we found the use of an in-vacuum mirror for coupling the Zeeman slowing laser beam in the chamber to be an efficient alternative to the commonly used heated viewports.

Besides a Zeeman slower, our apparatus features a 2D-MOT operated as a deflection and recollimation stage. While adding little complexity, this component is crucial to keeping a high fraction of the flux exiting the Zeeman slower capturable by the subsequent 3D trap. This is in particular relevant in setups, like ours, where the main experimental chamber needs to have a large physical volume due to other design constraints, at the expense of little loss of flux.

Our study involves a detailed characterization of the atomic beam emerging from a microchannel nozzle. We confirm that systems aiming at very high fluxes operate at the onset of the opaque channel regime, where interatomic interactions inside the nozzle cannot be neglected and the atomic beam divergence increases with increasing flux. This has crucial implications for the design of future devices aiming at even higher flux, for example for the detection of gravitational waves with atoms \cite{Loriani2019}.

Overall, our apparatus operates reliably at its maximum performance level with no other maintenance than that associated with laser systems at a level comparable to that achieved in other ytterbium \cite{Lee2015} or strontium \cite{Yang2015} setups. This constitutes a solid starting point for complex cold atoms experiments such as, but not limited to, high-performance atom interferometers.

%% file: paper.bbl
\begin{thebibliography}{64}%
\makeatletter
\providecommand \@ifxundefined [1]{%
 \@ifx{#1\undefined}
}%
\providecommand \@ifnum [1]{%
 \ifnum #1\expandafter \@firstoftwo
 \else \expandafter \@secondoftwo
 \fi
}%
\providecommand \@ifx [1]{%
 \ifx #1\expandafter \@firstoftwo
 \else \expandafter \@secondoftwo
 \fi
}%
\providecommand \natexlab [1]{#1}%
\providecommand \enquote  [1]{``#1''}%
\providecommand \bibnamefont  [1]{#1}%
\providecommand \bibfnamefont [1]{#1}%
\providecommand \citenamefont [1]{#1}%
\providecommand \href@noop [0]{\@secondoftwo}%
\providecommand \href [0]{\begingroup \@sanitize@url \@href}%
\providecommand \@href[1]{\@@startlink{#1}\@@href}%
\providecommand \@@href[1]{\endgroup#1\@@endlink}%
\providecommand \@sanitize@url [0]{\catcode `\\12\catcode `\$12\catcode
  `\&12\catcode `\#12\catcode `\^12\catcode `\_12\catcode `\%12\relax}%
\providecommand \@@startlink[1]{}%
\providecommand \@@endlink[0]{}%
\providecommand \url  [0]{\begingroup\@sanitize@url \@url }%
\providecommand \@url [1]{\endgroup\@href {#1}{\urlprefix }}%
\providecommand \urlprefix  [0]{URL }%
\providecommand \Eprint [0]{\href }%
\providecommand \doibase [0]{https://doi.org/}%
\providecommand \selectlanguage [0]{\@gobble}%
\providecommand \bibinfo  [0]{\@secondoftwo}%
\providecommand \bibfield  [0]{\@secondoftwo}%
\providecommand \translation [1]{[#1]}%
\providecommand \BibitemOpen [0]{}%
\providecommand \bibitemStop [0]{}%
\providecommand \bibitemNoStop [0]{.\EOS\space}%
\providecommand \EOS [0]{\spacefactor3000\relax}%
\providecommand \BibitemShut  [1]{\csname bibitem#1\endcsname}%
\let\auto@bib@innerbib\@empty
\bibitem [{\citenamefont {Gross}\ and\ \citenamefont
  {Bloch}(2017)}]{Gross2017}%
  \BibitemOpen
  \bibfield  {author} {\bibinfo {author} {\bibfnamefont {C.}~\bibnamefont
  {Gross}}\ and\ \bibinfo {author} {\bibfnamefont {I.}~\bibnamefont {Bloch}},\
  }\href {https://doi.org/10.1126/science.aal3837} {\bibfield  {journal}
  {\bibinfo  {journal} {Science}\ }\textbf {\bibinfo {volume} {357}},\ \bibinfo
  {pages} {995} (\bibinfo {year} {2017})}\BibitemShut {NoStop}%
\bibitem [{\citenamefont {Saffman}(2016)}]{Saffman2016}%
  \BibitemOpen
  \bibfield  {author} {\bibinfo {author} {\bibfnamefont {M.}~\bibnamefont
  {Saffman}},\ }\href {https://doi.org/10.1088/0953-4075/49/20/202001}
  {\bibfield  {journal} {\bibinfo  {journal} {J. Phys. B: At. Mol. Opt.}\
  }\textbf {\bibinfo {volume} {49}},\ \bibinfo {pages} {202001} (\bibinfo
  {year} {2016})}\BibitemShut {NoStop}%
\bibitem [{\citenamefont {Ludlow}\ \emph {et~al.}(2015)\citenamefont {Ludlow},
  \citenamefont {Boyd}, \citenamefont {Ye}, \citenamefont {Peik},\ and\
  \citenamefont {Schmidt}}]{Ludlow2015}%
  \BibitemOpen
  \bibfield  {author} {\bibinfo {author} {\bibfnamefont {A.~D.}\ \bibnamefont
  {Ludlow}}, \bibinfo {author} {\bibfnamefont {M.~M.}\ \bibnamefont {Boyd}},
  \bibinfo {author} {\bibfnamefont {J.}~\bibnamefont {Ye}}, \bibinfo {author}
  {\bibfnamefont {E.}~\bibnamefont {Peik}},\ and\ \bibinfo {author}
  {\bibfnamefont {P.~O.}\ \bibnamefont {Schmidt}},\ }\href
  {https://doi.org/10.1103/RevModPhys.87.637} {\bibfield  {journal} {\bibinfo
  {journal} {Rev. Mod. Phys.}\ }\textbf {\bibinfo {volume} {87}},\ \bibinfo
  {pages} {637} (\bibinfo {year} {2015})}\BibitemShut {NoStop}%
\bibitem [{\citenamefont {Rosi}\ \emph {et~al.}(2014)\citenamefont {Rosi},
  \citenamefont {Sorrentino}, \citenamefont {Cacciapuoti}, \citenamefont
  {Prevedelli},\ and\ \citenamefont {Tino}}]{Rosi2014}%
  \BibitemOpen
  \bibfield  {author} {\bibinfo {author} {\bibfnamefont {G.}~\bibnamefont
  {Rosi}}, \bibinfo {author} {\bibfnamefont {F.}~\bibnamefont {Sorrentino}},
  \bibinfo {author} {\bibfnamefont {L.}~\bibnamefont {Cacciapuoti}}, \bibinfo
  {author} {\bibfnamefont {M.}~\bibnamefont {Prevedelli}},\ and\ \bibinfo
  {author} {\bibfnamefont {G.}~\bibnamefont {Tino}},\ }\href
  {https://doi.org/10.1038/nature13433} {\bibfield  {journal} {\bibinfo
  {journal} {Nature}\ }\textbf {\bibinfo {volume} {510}},\ \bibinfo {pages}
  {518} (\bibinfo {year} {2014})}\BibitemShut {NoStop}%
\bibitem [{\citenamefont {Parker}\ \emph {et~al.}(2018)\citenamefont {Parker},
  \citenamefont {Yu}, \citenamefont {Zhong}, \citenamefont {Estey},\ and\
  \citenamefont {M{\"u}ller}}]{Parker2018}%
  \BibitemOpen
  \bibfield  {author} {\bibinfo {author} {\bibfnamefont {R.~H.}\ \bibnamefont
  {Parker}}, \bibinfo {author} {\bibfnamefont {C.}~\bibnamefont {Yu}}, \bibinfo
  {author} {\bibfnamefont {W.}~\bibnamefont {Zhong}}, \bibinfo {author}
  {\bibfnamefont {B.}~\bibnamefont {Estey}},\ and\ \bibinfo {author}
  {\bibfnamefont {H.}~\bibnamefont {M{\"u}ller}},\ }\href
  {https://doi.org/10.1126/science.aap7706} {\bibfield  {journal} {\bibinfo
  {journal} {Science}\ }\textbf {\bibinfo {volume} {360}},\ \bibinfo {pages}
  {191} (\bibinfo {year} {2018})}\BibitemShut {NoStop}%
\bibitem [{\citenamefont {Safronova}\ \emph {et~al.}(2018)\citenamefont
  {Safronova}, \citenamefont {Budker}, \citenamefont {DeMille}, \citenamefont
  {Kimball}, \citenamefont {Derevianko},\ and\ \citenamefont
  {Clark}}]{Safronova2018}%
  \BibitemOpen
  \bibfield  {author} {\bibinfo {author} {\bibfnamefont {M.~S.}\ \bibnamefont
  {Safronova}}, \bibinfo {author} {\bibfnamefont {D.}~\bibnamefont {Budker}},
  \bibinfo {author} {\bibfnamefont {D.}~\bibnamefont {DeMille}}, \bibinfo
  {author} {\bibfnamefont {D.~F.~J.}\ \bibnamefont {Kimball}}, \bibinfo
  {author} {\bibfnamefont {A.}~\bibnamefont {Derevianko}},\ and\ \bibinfo
  {author} {\bibfnamefont {C.~W.}\ \bibnamefont {Clark}},\ }\href
  {https://doi.org/10.1103/RevModPhys.90.025008} {\bibfield  {journal}
  {\bibinfo  {journal} {Rev. Mod. Phys.}\ }\textbf {\bibinfo {volume} {90}},\
  \bibinfo {pages} {025008} (\bibinfo {year} {2018})}\BibitemShut {NoStop}%
\bibitem [{\citenamefont {Dutta}\ \emph {et~al.}(2016)\citenamefont {Dutta},
  \citenamefont {Savoie}, \citenamefont {Fang}, \citenamefont {Venon},
  \citenamefont {Garrido~Alzar}, \citenamefont {Geiger},\ and\ \citenamefont
  {Landragin}}]{Dutta2016}%
  \BibitemOpen
  \bibfield  {author} {\bibinfo {author} {\bibfnamefont {I.}~\bibnamefont
  {Dutta}}, \bibinfo {author} {\bibfnamefont {D.}~\bibnamefont {Savoie}},
  \bibinfo {author} {\bibfnamefont {B.}~\bibnamefont {Fang}}, \bibinfo {author}
  {\bibfnamefont {B.}~\bibnamefont {Venon}}, \bibinfo {author} {\bibfnamefont
  {C.~L.}\ \bibnamefont {Garrido~Alzar}}, \bibinfo {author} {\bibfnamefont
  {R.}~\bibnamefont {Geiger}},\ and\ \bibinfo {author} {\bibfnamefont
  {A.}~\bibnamefont {Landragin}},\ }\href
  {https://doi.org/10.1103/PhysRevLett.116.183003} {\bibfield  {journal}
  {\bibinfo  {journal} {Phys. Rev. Lett.}\ }\textbf {\bibinfo {volume} {116}},\
  \bibinfo {eid} {183003} (\bibinfo {year} {2016})}\BibitemShut {NoStop}%
\bibitem [{\citenamefont {Freier}\ \emph {et~al.}(2016)\citenamefont {Freier},
  \citenamefont {Hauth}, \citenamefont {Schkolnik}, \citenamefont {Leykauf},
  \citenamefont {Schilling}, \citenamefont {Wziontek}, \citenamefont
  {Scherneck}, \citenamefont {M{\"u}ller},\ and\ \citenamefont
  {Peters}}]{Freier2016}%
  \BibitemOpen
  \bibfield  {author} {\bibinfo {author} {\bibfnamefont {C.}~\bibnamefont
  {Freier}}, \bibinfo {author} {\bibfnamefont {M.}~\bibnamefont {Hauth}},
  \bibinfo {author} {\bibfnamefont {V.}~\bibnamefont {Schkolnik}}, \bibinfo
  {author} {\bibfnamefont {B.}~\bibnamefont {Leykauf}}, \bibinfo {author}
  {\bibfnamefont {M.}~\bibnamefont {Schilling}}, \bibinfo {author}
  {\bibfnamefont {H.}~\bibnamefont {Wziontek}}, \bibinfo {author}
  {\bibfnamefont {H.-G.}\ \bibnamefont {Scherneck}}, \bibinfo {author}
  {\bibfnamefont {J.}~\bibnamefont {M{\"u}ller}},\ and\ \bibinfo {author}
  {\bibfnamefont {A.}~\bibnamefont {Peters}},\ }\href
  {https://doi.org/10.1088/1742-6596/723/1/012050} {\bibfield  {journal}
  {\bibinfo  {journal} {J. Phys: Conf. Ser.}\ }\textbf {\bibinfo {volume}
  {723}},\ \bibinfo {pages} {012050} (\bibinfo {year} {2016})}\BibitemShut
  {NoStop}%
\bibitem [{\citenamefont {Hill}\ \emph {et~al.}(2016)\citenamefont {Hill},
  \citenamefont {Hobson}, \citenamefont {Bowden}, \citenamefont {Bridge},
  \citenamefont {Donnellan}, \citenamefont {Curtis},\ and\ \citenamefont
  {Gill}}]{Hill2016}%
  \BibitemOpen
  \bibfield  {author} {\bibinfo {author} {\bibfnamefont {I.~R.}\ \bibnamefont
  {Hill}}, \bibinfo {author} {\bibfnamefont {R.}~\bibnamefont {Hobson}},
  \bibinfo {author} {\bibfnamefont {W.}~\bibnamefont {Bowden}}, \bibinfo
  {author} {\bibfnamefont {E.~M.}\ \bibnamefont {Bridge}}, \bibinfo {author}
  {\bibfnamefont {S.}~\bibnamefont {Donnellan}}, \bibinfo {author}
  {\bibfnamefont {E.~A.}\ \bibnamefont {Curtis}},\ and\ \bibinfo {author}
  {\bibfnamefont {P.}~\bibnamefont {Gill}},\ }\href
  {https://doi.org/10.1088/1742-6596/723/1/012019} {\bibfield  {journal}
  {\bibinfo  {journal} {J. Phys: Conf. Ser.}\ }\textbf {\bibinfo {volume}
  {723}},\ \bibinfo {pages} {012019} (\bibinfo {year} {2016})}\BibitemShut
  {NoStop}%
\bibitem [{\citenamefont {Grotti}\ \emph {et~al.}(2018)\citenamefont {Grotti},
  \citenamefont {Koller}, \citenamefont {Vogt}, \citenamefont {H{\"a}fner},
  \citenamefont {Sterr}, \citenamefont {Lisdat}, \citenamefont {Denker},
  \citenamefont {Voigt}, \citenamefont {Timmen}, \citenamefont {Rolland},
  \citenamefont {Baynes}, \citenamefont {Margolis}, \citenamefont {Zampaolo},
  \citenamefont {Thoumany}, \citenamefont {Pizzocaro}, \citenamefont {Rauf},
  \citenamefont {Bregolin}, \citenamefont {Tampellini}, \citenamefont
  {Barbiera}, \citenamefont {Zucco}, \citenamefont {Costanzo}, \citenamefont
  {Clivati}, \citenamefont {Levi},\ and\ \citenamefont
  {Calonico}}]{Grotti2018}%
  \BibitemOpen
  \bibfield  {author} {\bibinfo {author} {\bibfnamefont {J.}~\bibnamefont
  {Grotti}}, \bibinfo {author} {\bibfnamefont {S.}~\bibnamefont {Koller}},
  \bibinfo {author} {\bibfnamefont {S.}~\bibnamefont {Vogt}}, \bibinfo {author}
  {\bibfnamefont {S.}~\bibnamefont {H{\"a}fner}}, \bibinfo {author}
  {\bibfnamefont {U.}~\bibnamefont {Sterr}}, \bibinfo {author} {\bibfnamefont
  {C.}~\bibnamefont {Lisdat}}, \bibinfo {author} {\bibfnamefont
  {H.}~\bibnamefont {Denker}}, \bibinfo {author} {\bibfnamefont
  {C.}~\bibnamefont {Voigt}}, \bibinfo {author} {\bibfnamefont
  {L.}~\bibnamefont {Timmen}}, \bibinfo {author} {\bibfnamefont
  {A.}~\bibnamefont {Rolland}}, \bibinfo {author} {\bibfnamefont
  {F.}~\bibnamefont {Baynes}}, \bibinfo {author} {\bibfnamefont
  {H.}~\bibnamefont {Margolis}}, \bibinfo {author} {\bibfnamefont
  {M.}~\bibnamefont {Zampaolo}}, \bibinfo {author} {\bibfnamefont
  {P.}~\bibnamefont {Thoumany}}, \bibinfo {author} {\bibfnamefont
  {M.}~\bibnamefont {Pizzocaro}}, \bibinfo {author} {\bibfnamefont
  {B.}~\bibnamefont {Rauf}}, \bibinfo {author} {\bibfnamefont {F.}~\bibnamefont
  {Bregolin}}, \bibinfo {author} {\bibfnamefont {A.}~\bibnamefont
  {Tampellini}}, \bibinfo {author} {\bibfnamefont {P.}~\bibnamefont
  {Barbiera}}, \bibinfo {author} {\bibfnamefont {M.}~\bibnamefont {Zucco}},
  \bibinfo {author} {\bibfnamefont {G.}~\bibnamefont {Costanzo}}, \bibinfo
  {author} {\bibfnamefont {C.}~\bibnamefont {Clivati}}, \bibinfo {author}
  {\bibfnamefont {F.}~\bibnamefont {Levi}},\ and\ \bibinfo {author}
  {\bibfnamefont {D.}~\bibnamefont {Calonico}},\ }\href
  {https://doi.org/10.1038/s41567-017-0042-3} {\bibfield  {journal} {\bibinfo
  {journal} {Nat. Phys.}\ }\textbf {\bibinfo {volume} {14}},\ \bibinfo {pages}
  {437} (\bibinfo {year} {2018})}\BibitemShut {NoStop}%
\bibitem [{\citenamefont {Carraz}\ \emph {et~al.}(2009)\citenamefont {Carraz},
  \citenamefont {Lienhart}, \citenamefont {Charri{\`e}re}, \citenamefont
  {Cadoret}, \citenamefont {Zahzam}, \citenamefont {Bidel},\ and\ \citenamefont
  {Bresson}}]{Carraz2009}%
  \BibitemOpen
  \bibfield  {author} {\bibinfo {author} {\bibfnamefont {O.}~\bibnamefont
  {Carraz}}, \bibinfo {author} {\bibfnamefont {F.}~\bibnamefont {Lienhart}},
  \bibinfo {author} {\bibfnamefont {R.}~\bibnamefont {Charri{\`e}re}}, \bibinfo
  {author} {\bibfnamefont {M.}~\bibnamefont {Cadoret}}, \bibinfo {author}
  {\bibfnamefont {N.}~\bibnamefont {Zahzam}}, \bibinfo {author} {\bibfnamefont
  {Y.}~\bibnamefont {Bidel}},\ and\ \bibinfo {author} {\bibfnamefont
  {A.}~\bibnamefont {Bresson}},\ }\href
  {https://doi.org/10.1007/s00340-009-3675-9} {\bibfield  {journal} {\bibinfo
  {journal} {Appl. Phys. B}\ }\textbf {\bibinfo {volume} {97}},\ \bibinfo
  {pages} {405} (\bibinfo {year} {2009})}\BibitemShut {NoStop}%
\bibitem [{\citenamefont {Grosse}\ \emph {et~al.}(2016)\citenamefont {Grosse},
  \citenamefont {Seidel}, \citenamefont {Becker}, \citenamefont {Lachmann},
  \citenamefont {Scharringhausen}, \citenamefont {Braxmaier},\ and\
  \citenamefont {Rasel}}]{Grosse2016}%
  \BibitemOpen
  \bibfield  {author} {\bibinfo {author} {\bibfnamefont {J.}~\bibnamefont
  {Grosse}}, \bibinfo {author} {\bibfnamefont {S.~T.}\ \bibnamefont {Seidel}},
  \bibinfo {author} {\bibfnamefont {D.}~\bibnamefont {Becker}}, \bibinfo
  {author} {\bibfnamefont {M.~D.}\ \bibnamefont {Lachmann}}, \bibinfo {author}
  {\bibfnamefont {M.}~\bibnamefont {Scharringhausen}}, \bibinfo {author}
  {\bibfnamefont {C.}~\bibnamefont {Braxmaier}},\ and\ \bibinfo {author}
  {\bibfnamefont {E.~M.}\ \bibnamefont {Rasel}},\ }\href
  {https://doi.org/10.1116/1.4947583} {\bibfield  {journal} {\bibinfo
  {journal} {J. Vac. Sci. Technol. A}\ }\textbf {\bibinfo {volume} {34}},\
  \bibinfo {pages} {031606} (\bibinfo {year} {2016})}\BibitemShut {NoStop}%
\bibitem [{\citenamefont {Canuel}\ \emph {et~al.}(2018)\citenamefont {Canuel},
  \citenamefont {Bertoldi}, \citenamefont {Amand}, \citenamefont {Di~Borgo},
  \citenamefont {Chantrait}, \citenamefont {Danquigny}, \citenamefont
  {{\'A}lvarez}, \citenamefont {Fang}, \citenamefont {Freise}, \citenamefont
  {Geiger} \emph {et~al.}}]{Canuel2018}%
  \BibitemOpen
  \bibfield  {author} {\bibinfo {author} {\bibfnamefont {B.}~\bibnamefont
  {Canuel}}, \bibinfo {author} {\bibfnamefont {A.}~\bibnamefont {Bertoldi}},
  \bibinfo {author} {\bibfnamefont {L.}~\bibnamefont {Amand}}, \bibinfo
  {author} {\bibfnamefont {E.~P.}\ \bibnamefont {Di~Borgo}}, \bibinfo {author}
  {\bibfnamefont {T.}~\bibnamefont {Chantrait}}, \bibinfo {author}
  {\bibfnamefont {C.}~\bibnamefont {Danquigny}}, \bibinfo {author}
  {\bibfnamefont {M.~D.}\ \bibnamefont {{\'A}lvarez}}, \bibinfo {author}
  {\bibfnamefont {B.}~\bibnamefont {Fang}}, \bibinfo {author} {\bibfnamefont
  {A.}~\bibnamefont {Freise}}, \bibinfo {author} {\bibfnamefont
  {R.}~\bibnamefont {Geiger}}, \emph {et~al.},\ }\href
  {https://doi.org/10.1038/s41598-018-32165-z} {\bibfield  {journal} {\bibinfo
  {journal} {Sci. Rep.}\ }\textbf {\bibinfo {volume} {8}},\ \bibinfo {pages}
  {1} (\bibinfo {year} {2018})}\BibitemShut {NoStop}%
\bibitem [{\citenamefont {Coleman}(2018)}]{Coleman2018}%
  \BibitemOpen
  \bibfield  {author} {\bibinfo {author} {\bibfnamefont {J.}~\bibnamefont
  {Coleman}},\ }in\ \href {https://doi.org/10.22323/1.340.0021} {\emph
  {\bibinfo {booktitle} {Proceedings of the 39th International Conference on
  High-Energy Physics (ICHEP2018)}}}\ (\bibinfo {address} {Seoul, Korea},\
  \bibinfo {year} {2018})\BibitemShut {NoStop}%
\bibitem [{\citenamefont {Zhan}\ \emph {et~al.}(2019)\citenamefont {Zhan},
  \citenamefont {Wang}, \citenamefont {Ni}, \citenamefont {Gao}, \citenamefont
  {Wang}, \citenamefont {He}, \citenamefont {Li}, \citenamefont {Zhou},
  \citenamefont {Chen}, \citenamefont {Zhong}, \citenamefont {Tang} \emph
  {et~al.}}]{Zhan2019}%
  \BibitemOpen
  \bibfield  {author} {\bibinfo {author} {\bibfnamefont {M.-S.}\ \bibnamefont
  {Zhan}}, \bibinfo {author} {\bibfnamefont {J.}~\bibnamefont {Wang}}, \bibinfo
  {author} {\bibfnamefont {W.-T.}\ \bibnamefont {Ni}}, \bibinfo {author}
  {\bibfnamefont {D.-F.}\ \bibnamefont {Gao}}, \bibinfo {author} {\bibfnamefont
  {G.}~\bibnamefont {Wang}}, \bibinfo {author} {\bibfnamefont {L.-X.}\
  \bibnamefont {He}}, \bibinfo {author} {\bibfnamefont {R.-B.}\ \bibnamefont
  {Li}}, \bibinfo {author} {\bibfnamefont {L.}~\bibnamefont {Zhou}}, \bibinfo
  {author} {\bibfnamefont {X.}~\bibnamefont {Chen}}, \bibinfo {author}
  {\bibfnamefont {J.-Q.}\ \bibnamefont {Zhong}}, \bibinfo {author}
  {\bibfnamefont {B.}~\bibnamefont {Tang}}, \emph {et~al.},\ }\href
  {https://doi.org/10.1142/S0218271819400054} {\bibfield  {journal} {\bibinfo
  {journal} {Int. J. Mod. Phys. D}\ ,\ \bibinfo {pages} {1940005}} (\bibinfo
  {year} {2019})}\BibitemShut {NoStop}%
\bibitem [{\citenamefont {Barrett}\ \emph {et~al.}(2016)\citenamefont
  {Barrett}, \citenamefont {Antoni-Micollier}, \citenamefont {Chichet},
  \citenamefont {Battelier}, \citenamefont {L{\'e}v{\`e}que}, \citenamefont
  {Landragin},\ and\ \citenamefont {Bouyer}}]{Barrett2016}%
  \BibitemOpen
  \bibfield  {author} {\bibinfo {author} {\bibfnamefont {B.}~\bibnamefont
  {Barrett}}, \bibinfo {author} {\bibfnamefont {L.}~\bibnamefont
  {Antoni-Micollier}}, \bibinfo {author} {\bibfnamefont {L.}~\bibnamefont
  {Chichet}}, \bibinfo {author} {\bibfnamefont {B.}~\bibnamefont {Battelier}},
  \bibinfo {author} {\bibfnamefont {T.}~\bibnamefont {L{\'e}v{\`e}que}},
  \bibinfo {author} {\bibfnamefont {A.}~\bibnamefont {Landragin}},\ and\
  \bibinfo {author} {\bibfnamefont {P.}~\bibnamefont {Bouyer}},\ }\href
  {https://doi.org/10.1038/ncomms13786} {\bibfield  {journal} {\bibinfo
  {journal} {Nat. Commun.}\ }\textbf {\bibinfo {volume} {7}},\ \bibinfo {pages}
  {13786} (\bibinfo {year} {2016})}\BibitemShut {NoStop}%
\bibitem [{\citenamefont {Becker}\ \emph {et~al.}(2018)\citenamefont {Becker},
  \citenamefont {Lachmann}, \citenamefont {Seidel}, \citenamefont {Ahlers},
  \citenamefont {Dinkelaker}, \citenamefont {Grosse}, \citenamefont {Hellmig},
  \citenamefont {M{\"u}ntinga}, \citenamefont {Schkolnik}, \citenamefont
  {Wendrich} \emph {et~al.}}]{Becker2018}%
  \BibitemOpen
  \bibfield  {author} {\bibinfo {author} {\bibfnamefont {D.}~\bibnamefont
  {Becker}}, \bibinfo {author} {\bibfnamefont {M.~D.}\ \bibnamefont
  {Lachmann}}, \bibinfo {author} {\bibfnamefont {S.~T.}\ \bibnamefont
  {Seidel}}, \bibinfo {author} {\bibfnamefont {H.}~\bibnamefont {Ahlers}},
  \bibinfo {author} {\bibfnamefont {A.~N.}\ \bibnamefont {Dinkelaker}},
  \bibinfo {author} {\bibfnamefont {J.}~\bibnamefont {Grosse}}, \bibinfo
  {author} {\bibfnamefont {O.}~\bibnamefont {Hellmig}}, \bibinfo {author}
  {\bibfnamefont {H.}~\bibnamefont {M{\"u}ntinga}}, \bibinfo {author}
  {\bibfnamefont {V.}~\bibnamefont {Schkolnik}}, \bibinfo {author}
  {\bibfnamefont {T.}~\bibnamefont {Wendrich}}, \emph {et~al.},\ }\href
  {https://doi.org/10.1038/s41586-018-0605-1} {\bibfield  {journal} {\bibinfo
  {journal} {Nature}\ }\textbf {\bibinfo {volume} {562}},\ \bibinfo {pages}
  {391} (\bibinfo {year} {2018})}\BibitemShut {NoStop}%
\bibitem [{\citenamefont {Liu}\ \emph {et~al.}(2018)\citenamefont {Liu},
  \citenamefont {L{\"u}}, \citenamefont {Chen}, \citenamefont {Li},
  \citenamefont {Qu}, \citenamefont {Wang}, \citenamefont {Li}, \citenamefont
  {Ren}, \citenamefont {Dong}, \citenamefont {Zhao} \emph {et~al.}}]{Liu2018}%
  \BibitemOpen
  \bibfield  {author} {\bibinfo {author} {\bibfnamefont {L.}~\bibnamefont
  {Liu}}, \bibinfo {author} {\bibfnamefont {D.-S.}\ \bibnamefont {L{\"u}}},
  \bibinfo {author} {\bibfnamefont {W.-B.}\ \bibnamefont {Chen}}, \bibinfo
  {author} {\bibfnamefont {T.}~\bibnamefont {Li}}, \bibinfo {author}
  {\bibfnamefont {Q.-Z.}\ \bibnamefont {Qu}}, \bibinfo {author} {\bibfnamefont
  {B.}~\bibnamefont {Wang}}, \bibinfo {author} {\bibfnamefont {L.}~\bibnamefont
  {Li}}, \bibinfo {author} {\bibfnamefont {W.}~\bibnamefont {Ren}}, \bibinfo
  {author} {\bibfnamefont {Z.-R.}\ \bibnamefont {Dong}}, \bibinfo {author}
  {\bibfnamefont {J.-B.}\ \bibnamefont {Zhao}}, \emph {et~al.},\ }\href
  {https://doi.org/10.1038/s41467-018-05219-z} {\bibfield  {journal} {\bibinfo
  {journal} {Nat. Commun.}\ }\textbf {\bibinfo {volume} {9}},\ \bibinfo {pages}
  {2760} (\bibinfo {year} {2018})}\BibitemShut {NoStop}%
\bibitem [{\citenamefont {Frye}\ \emph {et~al.}(2019)\citenamefont {Frye},
  \citenamefont {Abend}, \citenamefont {Bartosch}, \citenamefont {Bawamia},
  \citenamefont {Becker}, \citenamefont {Blume}, \citenamefont {Braxmaier},
  \citenamefont {Chiow}, \citenamefont {Efremov}, \citenamefont {Ertmer} \emph
  {et~al.}}]{Frye2019}%
  \BibitemOpen
  \bibfield  {author} {\bibinfo {author} {\bibfnamefont {K.}~\bibnamefont
  {Frye}}, \bibinfo {author} {\bibfnamefont {S.}~\bibnamefont {Abend}},
  \bibinfo {author} {\bibfnamefont {W.}~\bibnamefont {Bartosch}}, \bibinfo
  {author} {\bibfnamefont {A.}~\bibnamefont {Bawamia}}, \bibinfo {author}
  {\bibfnamefont {D.}~\bibnamefont {Becker}}, \bibinfo {author} {\bibfnamefont
  {H.}~\bibnamefont {Blume}}, \bibinfo {author} {\bibfnamefont
  {C.}~\bibnamefont {Braxmaier}}, \bibinfo {author} {\bibfnamefont {S.-W.}\
  \bibnamefont {Chiow}}, \bibinfo {author} {\bibfnamefont {M.~A.}\ \bibnamefont
  {Efremov}}, \bibinfo {author} {\bibfnamefont {W.}~\bibnamefont {Ertmer}},
  \emph {et~al.},\ }\href@noop {} {\bibinfo {title} {The {B}ose-{E}instein
  {C}ondensate and {C}old {A}tom {L}aboratory}} (\bibinfo {year} {2019}),\
  \Eprint {https://arxiv.org/abs/1912.04849} {arXiv:1912.04849
  [physics.atom-ph]} \BibitemShut {NoStop}%
\bibitem [{\citenamefont {Aguilera}\ \emph {et~al.}(2014)\citenamefont
  {Aguilera}, \citenamefont {Ahlers}, \citenamefont {Battelier}, \citenamefont
  {Bawamia}, \citenamefont {Bertoldi}, \citenamefont {Bondarescu},
  \citenamefont {Bongs}, \citenamefont {Bouyer}, \citenamefont {Braxmaier},
  \citenamefont {Cacciapuoti} \emph {et~al.}}]{Aguilera2014}%
  \BibitemOpen
  \bibfield  {author} {\bibinfo {author} {\bibfnamefont {D.}~\bibnamefont
  {Aguilera}}, \bibinfo {author} {\bibfnamefont {H.}~\bibnamefont {Ahlers}},
  \bibinfo {author} {\bibfnamefont {B.}~\bibnamefont {Battelier}}, \bibinfo
  {author} {\bibfnamefont {A.}~\bibnamefont {Bawamia}}, \bibinfo {author}
  {\bibfnamefont {A.}~\bibnamefont {Bertoldi}}, \bibinfo {author}
  {\bibfnamefont {R.}~\bibnamefont {Bondarescu}}, \bibinfo {author}
  {\bibfnamefont {K.}~\bibnamefont {Bongs}}, \bibinfo {author} {\bibfnamefont
  {P.}~\bibnamefont {Bouyer}}, \bibinfo {author} {\bibfnamefont
  {C.}~\bibnamefont {Braxmaier}}, \bibinfo {author} {\bibfnamefont
  {L.}~\bibnamefont {Cacciapuoti}}, \emph {et~al.},\ }\href
  {https://doi.org/10.1088/0264-9381/31/11/115010} {\bibfield  {journal}
  {\bibinfo  {journal} {Class. Quant. Grav.}\ }\textbf {\bibinfo {volume}
  {31}},\ \bibinfo {pages} {115010} (\bibinfo {year} {2014})}\BibitemShut
  {NoStop}%
\bibitem [{\citenamefont {Gao}\ \emph {et~al.}(2018)\citenamefont {Gao},
  \citenamefont {Wang},\ and\ \citenamefont {Zhan}}]{Gao2018}%
  \BibitemOpen
  \bibfield  {author} {\bibinfo {author} {\bibfnamefont {D.-F.}\ \bibnamefont
  {Gao}}, \bibinfo {author} {\bibfnamefont {J.}~\bibnamefont {Wang}},\ and\
  \bibinfo {author} {\bibfnamefont {M.-S.}\ \bibnamefont {Zhan}},\ }\href
  {https://doi.org/10.1088/0253-6102/69/1/37} {\bibfield  {journal} {\bibinfo
  {journal} {Commun. Theor. Phys.}\ }\textbf {\bibinfo {volume} {69}},\
  \bibinfo {pages} {37} (\bibinfo {year} {2018})}\BibitemShut {NoStop}%
\bibitem [{\citenamefont {Kolkowitz}\ \emph {et~al.}(2016)\citenamefont
  {Kolkowitz}, \citenamefont {Pikovski}, \citenamefont {Langellier},
  \citenamefont {Lukin}, \citenamefont {Walsworth},\ and\ \citenamefont
  {Ye}}]{Kolkowitz2016}%
  \BibitemOpen
  \bibfield  {author} {\bibinfo {author} {\bibfnamefont {S.}~\bibnamefont
  {Kolkowitz}}, \bibinfo {author} {\bibfnamefont {I.}~\bibnamefont {Pikovski}},
  \bibinfo {author} {\bibfnamefont {N.}~\bibnamefont {Langellier}}, \bibinfo
  {author} {\bibfnamefont {M.~D.}\ \bibnamefont {Lukin}}, \bibinfo {author}
  {\bibfnamefont {R.~L.}\ \bibnamefont {Walsworth}},\ and\ \bibinfo {author}
  {\bibfnamefont {J.}~\bibnamefont {Ye}},\ }\href
  {https://doi.org/10.1103/PhysRevD.94.124043} {\bibfield  {journal} {\bibinfo
  {journal} {Phys. Rev. D}\ }\textbf {\bibinfo {volume} {94}},\ \bibinfo
  {pages} {124043} (\bibinfo {year} {2016})}\BibitemShut {NoStop}%
\bibitem [{\citenamefont {Hogan}\ and\ \citenamefont
  {Kasevich}(2016)}]{Hogan2016}%
  \BibitemOpen
  \bibfield  {author} {\bibinfo {author} {\bibfnamefont {J.~M.}\ \bibnamefont
  {Hogan}}\ and\ \bibinfo {author} {\bibfnamefont {M.~A.}\ \bibnamefont
  {Kasevich}},\ }\href {https://doi.org/10.1103/PhysRevA.94.033632} {\bibfield
  {journal} {\bibinfo  {journal} {Phys. Rev. A}\ }\textbf {\bibinfo {volume}
  {94}},\ \bibinfo {pages} {033632} (\bibinfo {year} {2016})}\BibitemShut
  {NoStop}%
\bibitem [{\citenamefont {Hogan}\ \emph {et~al.}(2008)\citenamefont {Hogan},
  \citenamefont {Johnson},\ and\ \citenamefont {Kasevich}}]{Hogan2008}%
  \BibitemOpen
  \bibfield  {author} {\bibinfo {author} {\bibfnamefont {J.~M.}\ \bibnamefont
  {Hogan}}, \bibinfo {author} {\bibfnamefont {D.}~\bibnamefont {Johnson}},\
  and\ \bibinfo {author} {\bibfnamefont {M.~A.}\ \bibnamefont {Kasevich}},\
  }\href@noop {} {\bibinfo {title} {Light-pulse atom interferometry}} (\bibinfo
  {year} {2008}),\ \Eprint {https://arxiv.org/abs/0806.3261} {arXiv:0806.3261
  [physics.atom-ph]} \BibitemShut {NoStop}%
\bibitem [{\citenamefont {Savoie}\ \emph {et~al.}(2018)\citenamefont {Savoie},
  \citenamefont {Altorio}, \citenamefont {Fang}, \citenamefont {Sidorenkov},
  \citenamefont {Geiger},\ and\ \citenamefont {Landragin}}]{Savoie2018}%
  \BibitemOpen
  \bibfield  {author} {\bibinfo {author} {\bibfnamefont {D.}~\bibnamefont
  {Savoie}}, \bibinfo {author} {\bibfnamefont {M.}~\bibnamefont {Altorio}},
  \bibinfo {author} {\bibfnamefont {B.}~\bibnamefont {Fang}}, \bibinfo {author}
  {\bibfnamefont {L.}~\bibnamefont {Sidorenkov}}, \bibinfo {author}
  {\bibfnamefont {R.}~\bibnamefont {Geiger}},\ and\ \bibinfo {author}
  {\bibfnamefont {A.}~\bibnamefont {Landragin}},\ }\href
  {https://doi.org/10.1126/sciadv.aau7948} {\bibfield  {journal} {\bibinfo
  {journal} {Sci. Adv.}\ }\textbf {\bibinfo {volume} {4}},\ \bibinfo {pages}
  {eaau7948} (\bibinfo {year} {2018})}\BibitemShut {NoStop}%
\bibitem [{\citenamefont {Canuel}\ \emph {et~al.}(2006)\citenamefont {Canuel},
  \citenamefont {Leduc}, \citenamefont {Holleville}, \citenamefont {Gauguet},
  \citenamefont {Fils}, \citenamefont {Virdis}, \citenamefont {Clairon},
  \citenamefont {Dimarcq}, \citenamefont {Bord\'e}, \citenamefont {Landragin},\
  and\ \citenamefont {Bouyer}}]{Canuel2006}%
  \BibitemOpen
  \bibfield  {author} {\bibinfo {author} {\bibfnamefont {B.}~\bibnamefont
  {Canuel}}, \bibinfo {author} {\bibfnamefont {F.}~\bibnamefont {Leduc}},
  \bibinfo {author} {\bibfnamefont {D.}~\bibnamefont {Holleville}}, \bibinfo
  {author} {\bibfnamefont {A.}~\bibnamefont {Gauguet}}, \bibinfo {author}
  {\bibfnamefont {J.}~\bibnamefont {Fils}}, \bibinfo {author} {\bibfnamefont
  {A.}~\bibnamefont {Virdis}}, \bibinfo {author} {\bibfnamefont
  {A.}~\bibnamefont {Clairon}}, \bibinfo {author} {\bibfnamefont
  {N.}~\bibnamefont {Dimarcq}}, \bibinfo {author} {\bibfnamefont {C.~J.}\
  \bibnamefont {Bord\'e}}, \bibinfo {author} {\bibfnamefont {A.}~\bibnamefont
  {Landragin}},\ and\ \bibinfo {author} {\bibfnamefont {P.}~\bibnamefont
  {Bouyer}},\ }\href {https://doi.org/10.1103/PhysRevLett.97.010402} {\bibfield
   {journal} {\bibinfo  {journal} {Phys. Rev. Lett.}\ }\textbf {\bibinfo
  {volume} {97}},\ \bibinfo {pages} {010402} (\bibinfo {year}
  {2006})}\BibitemShut {NoStop}%
\bibitem [{\citenamefont {Garrido~Alzar}(2019)}]{Garrido2019}%
  \BibitemOpen
  \bibfield  {author} {\bibinfo {author} {\bibfnamefont {C.~L.}\ \bibnamefont
  {Garrido~Alzar}},\ }\href {https://doi.org/10.1116/1.5120348} {\bibfield
  {journal} {\bibinfo  {journal} {AVS Quantum Sci.}\ }\textbf {\bibinfo
  {volume} {1}},\ \bibinfo {pages} {014702} (\bibinfo {year}
  {2019})}\BibitemShut {NoStop}%
\bibitem [{\citenamefont {Hu}\ \emph {et~al.}(2017)\citenamefont {Hu},
  \citenamefont {Poli}, \citenamefont {Salvi},\ and\ \citenamefont
  {Tino}}]{Hu2017}%
  \BibitemOpen
  \bibfield  {author} {\bibinfo {author} {\bibfnamefont {L.}~\bibnamefont
  {Hu}}, \bibinfo {author} {\bibfnamefont {N.}~\bibnamefont {Poli}}, \bibinfo
  {author} {\bibfnamefont {L.}~\bibnamefont {Salvi}},\ and\ \bibinfo {author}
  {\bibfnamefont {G.~M.}\ \bibnamefont {Tino}},\ }\href
  {https://doi.org/10.1103/PhysRevLett.119.263601} {\bibfield  {journal}
  {\bibinfo  {journal} {Phys. Rev. Lett.}\ }\textbf {\bibinfo {volume} {119}},\
  \bibinfo {pages} {263601} (\bibinfo {year} {2017})}\BibitemShut {NoStop}%
\bibitem [{\citenamefont {Plotkin-Swing}\ \emph {et~al.}(2018)\citenamefont
  {Plotkin-Swing}, \citenamefont {Gochnauer}, \citenamefont {McAlpine},
  \citenamefont {Cooper}, \citenamefont {Jamison},\ and\ \citenamefont
  {Gupta}}]{Plotkin-Swing2018}%
  \BibitemOpen
  \bibfield  {author} {\bibinfo {author} {\bibfnamefont {B.}~\bibnamefont
  {Plotkin-Swing}}, \bibinfo {author} {\bibfnamefont {D.}~\bibnamefont
  {Gochnauer}}, \bibinfo {author} {\bibfnamefont {K.~E.}\ \bibnamefont
  {McAlpine}}, \bibinfo {author} {\bibfnamefont {E.~S.}\ \bibnamefont
  {Cooper}}, \bibinfo {author} {\bibfnamefont {A.~O.}\ \bibnamefont
  {Jamison}},\ and\ \bibinfo {author} {\bibfnamefont {S.}~\bibnamefont
  {Gupta}},\ }\href {https://doi.org/10.1103/PhysRevLett.121.133201} {\bibfield
   {journal} {\bibinfo  {journal} {Phys. Rev. Lett.}\ }\textbf {\bibinfo
  {volume} {121}},\ \bibinfo {pages} {133201} (\bibinfo {year}
  {2018})}\BibitemShut {NoStop}%
\bibitem [{\citenamefont {Rudolph}\ \emph {et~al.}(2020)\citenamefont
  {Rudolph}, \citenamefont {Wilkason}, \citenamefont {Nantel}, \citenamefont
  {Swan}, \citenamefont {Holland}, \citenamefont {Jiang}, \citenamefont
  {Garber}, \citenamefont {Carman},\ and\ \citenamefont {Hogan}}]{Rudolph2020}%
  \BibitemOpen
  \bibfield  {author} {\bibinfo {author} {\bibfnamefont {J.}~\bibnamefont
  {Rudolph}}, \bibinfo {author} {\bibfnamefont {T.}~\bibnamefont {Wilkason}},
  \bibinfo {author} {\bibfnamefont {M.}~\bibnamefont {Nantel}}, \bibinfo
  {author} {\bibfnamefont {H.}~\bibnamefont {Swan}}, \bibinfo {author}
  {\bibfnamefont {C.~M.}\ \bibnamefont {Holland}}, \bibinfo {author}
  {\bibfnamefont {Y.}~\bibnamefont {Jiang}}, \bibinfo {author} {\bibfnamefont
  {B.~E.}\ \bibnamefont {Garber}}, \bibinfo {author} {\bibfnamefont {S.~P.}\
  \bibnamefont {Carman}},\ and\ \bibinfo {author} {\bibfnamefont {J.~M.}\
  \bibnamefont {Hogan}},\ }\href
  {https://doi.org/10.1103/PhysRevLett.124.083604} {\bibfield  {journal}
  {\bibinfo  {journal} {Phys. Rev. Lett.}\ }\textbf {\bibinfo {volume} {124}},\
  \bibinfo {pages} {083604} (\bibinfo {year} {2020})}\BibitemShut {NoStop}%
\bibitem [{\citenamefont {Loriani}\ \emph {et~al.}(2019)\citenamefont
  {Loriani}, \citenamefont {Schlippert}, \citenamefont {Schubert},
  \citenamefont {Abend}, \citenamefont {Ahlers}, \citenamefont {Ertmer},
  \citenamefont {Rudolph}, \citenamefont {Hogan}, \citenamefont {Kasevich},
  \citenamefont {Rasel} \emph {et~al.}}]{Loriani2019}%
  \BibitemOpen
  \bibfield  {author} {\bibinfo {author} {\bibfnamefont {S.}~\bibnamefont
  {Loriani}}, \bibinfo {author} {\bibfnamefont {D.}~\bibnamefont {Schlippert}},
  \bibinfo {author} {\bibfnamefont {C.}~\bibnamefont {Schubert}}, \bibinfo
  {author} {\bibfnamefont {S.}~\bibnamefont {Abend}}, \bibinfo {author}
  {\bibfnamefont {H.}~\bibnamefont {Ahlers}}, \bibinfo {author} {\bibfnamefont
  {W.}~\bibnamefont {Ertmer}}, \bibinfo {author} {\bibfnamefont
  {J.}~\bibnamefont {Rudolph}}, \bibinfo {author} {\bibfnamefont
  {J.}~\bibnamefont {Hogan}}, \bibinfo {author} {\bibfnamefont
  {M.}~\bibnamefont {Kasevich}}, \bibinfo {author} {\bibfnamefont
  {E.}~\bibnamefont {Rasel}}, \emph {et~al.},\ }\href
  {https://doi.org/10.1088/1367-2630/ab22d0} {\bibfield  {journal} {\bibinfo
  {journal} {New J. Phys.}\ }\textbf {\bibinfo {volume} {21}},\ \bibinfo
  {pages} {063030} (\bibinfo {year} {2019})}\BibitemShut {NoStop}%
\bibitem [{\citenamefont {Hartwig}\ \emph {et~al.}(2015)\citenamefont
  {Hartwig}, \citenamefont {Abend}, \citenamefont {Schubert}, \citenamefont
  {Schlippert}, \citenamefont {Ahlers}, \citenamefont {Posso-Trujillo},
  \citenamefont {Gaaloul}, \citenamefont {Ertmer},\ and\ \citenamefont
  {Rasel}}]{Hartwig2015}%
  \BibitemOpen
  \bibfield  {author} {\bibinfo {author} {\bibfnamefont {J.}~\bibnamefont
  {Hartwig}}, \bibinfo {author} {\bibfnamefont {S.}~\bibnamefont {Abend}},
  \bibinfo {author} {\bibfnamefont {C.}~\bibnamefont {Schubert}}, \bibinfo
  {author} {\bibfnamefont {D.}~\bibnamefont {Schlippert}}, \bibinfo {author}
  {\bibfnamefont {H.}~\bibnamefont {Ahlers}}, \bibinfo {author} {\bibfnamefont
  {K.}~\bibnamefont {Posso-Trujillo}}, \bibinfo {author} {\bibfnamefont
  {N.}~\bibnamefont {Gaaloul}}, \bibinfo {author} {\bibfnamefont
  {W.}~\bibnamefont {Ertmer}},\ and\ \bibinfo {author} {\bibfnamefont {E.~M.}\
  \bibnamefont {Rasel}},\ }\href
  {https://doi.org/10.1088/1367-2630/17/3/035011} {\bibfield  {journal}
  {\bibinfo  {journal} {New J. Phys.}\ }\textbf {\bibinfo {volume} {17}},\
  \bibinfo {pages} {035011} (\bibinfo {year} {2015})}\BibitemShut {NoStop}%
\bibitem [{\citenamefont {Takasu}\ \emph {et~al.}(2003)\citenamefont {Takasu},
  \citenamefont {Maki}, \citenamefont {Komori}, \citenamefont {Takano},
  \citenamefont {Honda}, \citenamefont {Kumakura}, \citenamefont {Yabuzaki},\
  and\ \citenamefont {Takahashi}}]{Takasu2003}%
  \BibitemOpen
  \bibfield  {author} {\bibinfo {author} {\bibfnamefont {Y.}~\bibnamefont
  {Takasu}}, \bibinfo {author} {\bibfnamefont {K.}~\bibnamefont {Maki}},
  \bibinfo {author} {\bibfnamefont {K.}~\bibnamefont {Komori}}, \bibinfo
  {author} {\bibfnamefont {T.}~\bibnamefont {Takano}}, \bibinfo {author}
  {\bibfnamefont {K.}~\bibnamefont {Honda}}, \bibinfo {author} {\bibfnamefont
  {M.}~\bibnamefont {Kumakura}}, \bibinfo {author} {\bibfnamefont
  {T.}~\bibnamefont {Yabuzaki}},\ and\ \bibinfo {author} {\bibfnamefont
  {Y.}~\bibnamefont {Takahashi}},\ }\href
  {https://doi.org/10.1103/PhysRevLett.91.040404} {\bibfield  {journal}
  {\bibinfo  {journal} {Phys. Rev. Lett.}\ }\textbf {\bibinfo {volume} {91}},\
  \bibinfo {pages} {040404} (\bibinfo {year} {2003})}\BibitemShut {NoStop}%
\bibitem [{\citenamefont {D{\"o}rscher}\ \emph {et~al.}(2013)\citenamefont
  {D{\"o}rscher}, \citenamefont {Thobe}, \citenamefont {Hundt}, \citenamefont
  {Kochanke}, \citenamefont {Le~Targat}, \citenamefont {Windpassinger},
  \citenamefont {Becker},\ and\ \citenamefont {Sengstock}}]{Dorscher2013}%
  \BibitemOpen
  \bibfield  {author} {\bibinfo {author} {\bibfnamefont {S.}~\bibnamefont
  {D{\"o}rscher}}, \bibinfo {author} {\bibfnamefont {A.}~\bibnamefont {Thobe}},
  \bibinfo {author} {\bibfnamefont {B.}~\bibnamefont {Hundt}}, \bibinfo
  {author} {\bibfnamefont {A.}~\bibnamefont {Kochanke}}, \bibinfo {author}
  {\bibfnamefont {R.}~\bibnamefont {Le~Targat}}, \bibinfo {author}
  {\bibfnamefont {P.}~\bibnamefont {Windpassinger}}, \bibinfo {author}
  {\bibfnamefont {C.}~\bibnamefont {Becker}},\ and\ \bibinfo {author}
  {\bibfnamefont {K.}~\bibnamefont {Sengstock}},\ }\href
  {https://doi.org/10.1063/1.4802682} {\bibfield  {journal} {\bibinfo
  {journal} {Rev. Sci. Instrum.}\ }\textbf {\bibinfo {volume} {84}},\ \bibinfo
  {pages} {043109} (\bibinfo {year} {2013})}\BibitemShut {NoStop}%
\bibitem [{\citenamefont {Yang}\ \emph {et~al.}(2015)\citenamefont {Yang},
  \citenamefont {Pandey}, \citenamefont {Pramod}, \citenamefont {Leroux},
  \citenamefont {Kwong}, \citenamefont {Hajiyev}, \citenamefont {Chia},
  \citenamefont {Fang},\ and\ \citenamefont {Wilkowski}}]{Yang2015}%
  \BibitemOpen
  \bibfield  {author} {\bibinfo {author} {\bibfnamefont {T.}~\bibnamefont
  {Yang}}, \bibinfo {author} {\bibfnamefont {K.}~\bibnamefont {Pandey}},
  \bibinfo {author} {\bibfnamefont {M.~S.}\ \bibnamefont {Pramod}}, \bibinfo
  {author} {\bibfnamefont {F.}~\bibnamefont {Leroux}}, \bibinfo {author}
  {\bibfnamefont {C.~C.}\ \bibnamefont {Kwong}}, \bibinfo {author}
  {\bibfnamefont {E.}~\bibnamefont {Hajiyev}}, \bibinfo {author} {\bibfnamefont
  {Z.~Y.}\ \bibnamefont {Chia}}, \bibinfo {author} {\bibfnamefont
  {B.}~\bibnamefont {Fang}},\ and\ \bibinfo {author} {\bibfnamefont
  {D.}~\bibnamefont {Wilkowski}},\ }\href
  {https://doi.org/10.1140/epjd/e2015-60288-y} {\bibfield  {journal} {\bibinfo
  {journal} {Eur. Phys. J. D}\ }\textbf {\bibinfo {volume} {69}},\ \bibinfo
  {pages} {226} (\bibinfo {year} {2015})}\BibitemShut {NoStop}%
\bibitem [{\citenamefont {Lee}\ \emph {et~al.}(2015)\citenamefont {Lee},
  \citenamefont {Lee}, \citenamefont {Noh},\ and\ \citenamefont
  {Mun}}]{Lee2015}%
  \BibitemOpen
  \bibfield  {author} {\bibinfo {author} {\bibfnamefont {J.}~\bibnamefont
  {Lee}}, \bibinfo {author} {\bibfnamefont {J.~H.}\ \bibnamefont {Lee}},
  \bibinfo {author} {\bibfnamefont {J.}~\bibnamefont {Noh}},\ and\ \bibinfo
  {author} {\bibfnamefont {J.}~\bibnamefont {Mun}},\ }\href
  {https://doi.org/10.1103/PhysRevA.91.053405} {\bibfield  {journal} {\bibinfo
  {journal} {Phys. Rev. A}\ }\textbf {\bibinfo {volume} {91}},\ \bibinfo
  {pages} {053405} (\bibinfo {year} {2015})}\BibitemShut {NoStop}%
\bibitem [{\citenamefont {Senaratne}\ \emph {et~al.}(2015)\citenamefont
  {Senaratne}, \citenamefont {Rajagopal}, \citenamefont {Geiger}, \citenamefont
  {Fujiwara}, \citenamefont {Lebedev},\ and\ \citenamefont
  {Weld}}]{Senaratne2015}%
  \BibitemOpen
  \bibfield  {author} {\bibinfo {author} {\bibfnamefont {R.}~\bibnamefont
  {Senaratne}}, \bibinfo {author} {\bibfnamefont {S.~V.}\ \bibnamefont
  {Rajagopal}}, \bibinfo {author} {\bibfnamefont {Z.~A.}\ \bibnamefont
  {Geiger}}, \bibinfo {author} {\bibfnamefont {K.~M.}\ \bibnamefont
  {Fujiwara}}, \bibinfo {author} {\bibfnamefont {V.}~\bibnamefont {Lebedev}},\
  and\ \bibinfo {author} {\bibfnamefont {D.~M.}\ \bibnamefont {Weld}},\ }\href
  {https://doi.org/10.1063/1.4907401} {\bibfield  {journal} {\bibinfo
  {journal} {Rev. Sci. Instrum.}\ }\textbf {\bibinfo {volume} {86}},\ \bibinfo
  {pages} {023105} (\bibinfo {year} {2015})}\BibitemShut {NoStop}%
\bibitem [{\citenamefont {Deilamian}\ \emph {et~al.}(1993)\citenamefont
  {Deilamian}, \citenamefont {Gillaspy},\ and\ \citenamefont
  {Kelleher}}]{Deilamian1993}%
  \BibitemOpen
  \bibfield  {author} {\bibinfo {author} {\bibfnamefont {K.}~\bibnamefont
  {Deilamian}}, \bibinfo {author} {\bibfnamefont {J.~D.}\ \bibnamefont
  {Gillaspy}},\ and\ \bibinfo {author} {\bibfnamefont {D.~E.}\ \bibnamefont
  {Kelleher}},\ }\href {https://doi.org/10.1364/JOSAB.10.000789} {\bibfield
  {journal} {\bibinfo  {journal} {J. Opt. Soc. Am. B}\ }\textbf {\bibinfo
  {volume} {10}},\ \bibinfo {pages} {789} (\bibinfo {year} {1993})}\BibitemShut
  {NoStop}%
\bibitem [{\citenamefont {Das}\ \emph {et~al.}(2005)\citenamefont {Das},
  \citenamefont {Barthwal}, \citenamefont {Banerjee},\ and\ \citenamefont
  {Natarajan}}]{Das2005}%
  \BibitemOpen
  \bibfield  {author} {\bibinfo {author} {\bibfnamefont {D.}~\bibnamefont
  {Das}}, \bibinfo {author} {\bibfnamefont {S.}~\bibnamefont {Barthwal}},
  \bibinfo {author} {\bibfnamefont {A.}~\bibnamefont {Banerjee}},\ and\
  \bibinfo {author} {\bibfnamefont {V.}~\bibnamefont {Natarajan}},\ }\href
  {https://doi.org/10.1103/PhysRevA.72.032506} {\bibfield  {journal} {\bibinfo
  {journal} {Phys. Rev. A}\ }\textbf {\bibinfo {volume} {72}},\ \bibinfo
  {pages} {032506} (\bibinfo {year} {2005})}\BibitemShut {NoStop}%
\bibitem [{\citenamefont {Degl'Innocenti}(2014)}]{DeglInnocenti2014}%
  \BibitemOpen
  \bibfield  {author} {\bibinfo {author} {\bibfnamefont {E.~L.}\ \bibnamefont
  {Degl'Innocenti}},\ }\href {https://doi.org/10.1007/978-88-470-2808-1} {\emph
  {\bibinfo {title} {Atomic spectroscopy and radiative processes}}}\ (\bibinfo
  {publisher} {Springer},\ \bibinfo {address} {Milano},\ \bibinfo {year}
  {2014})\BibitemShut {NoStop}%
\bibitem [{\citenamefont {Metcalf}\ and\ \citenamefont {van~der
  Straten}(1999)}]{Metcalf1999}%
  \BibitemOpen
  \bibfield  {author} {\bibinfo {author} {\bibfnamefont {H.~J.}\ \bibnamefont
  {Metcalf}}\ and\ \bibinfo {author} {\bibfnamefont {P.}~\bibnamefont {van~der
  Straten}},\ }\href {https://doi.org/10.1007/978-1-4612-1470-0} {\emph
  {\bibinfo {title} {Laser Cooling and Trapping}}}\ (\bibinfo  {publisher}
  {Springer},\ \bibinfo {address} {New York},\ \bibinfo {year}
  {1999})\BibitemShut {NoStop}%
\bibitem [{\citenamefont {Giordmaine}\ and\ \citenamefont
  {Wang}(1960)}]{Giordmaine1960}%
  \BibitemOpen
  \bibfield  {author} {\bibinfo {author} {\bibfnamefont {J.}~\bibnamefont
  {Giordmaine}}\ and\ \bibinfo {author} {\bibfnamefont {T.}~\bibnamefont
  {Wang}},\ }\href {https://doi.org/10.1063/1.1735609} {\bibfield  {journal}
  {\bibinfo  {journal} {J. Appl. Phys.}\ }\textbf {\bibinfo {volume} {31}},\
  \bibinfo {pages} {463} (\bibinfo {year} {1960})}\BibitemShut {NoStop}%
\bibitem [{\citenamefont {Ramsey}(1956)}]{Ramsey1956}%
  \BibitemOpen
  \bibfield  {author} {\bibinfo {author} {\bibfnamefont {N.}~\bibnamefont
  {Ramsey}},\ }\href@noop {} {\emph {\bibinfo {title} {Molecular beams}}},\
  Vol.~\bibinfo {volume} {20}\ (\bibinfo  {publisher} {Oxford University
  Press},\ \bibinfo {year} {1956})\BibitemShut {NoStop}%
\bibitem [{\citenamefont {Lide}(1997)}]{CRCHandbook1997}%
  \BibitemOpen
  \bibfield  {author} {\bibinfo {author} {\bibfnamefont {D.~R.}\ \bibnamefont
  {Lide}},\ }\href@noop {} {\emph {\bibinfo {title} {Handbook of Chemistry and
  Physics}}},\ Vol.~\bibinfo {volume} {78}\ (\bibinfo  {publisher} {CRC
  press},\ \bibinfo {year} {1997})\BibitemShut {NoStop}%
\bibitem [{\citenamefont {Schioppo}\ \emph {et~al.}(2012)\citenamefont
  {Schioppo}, \citenamefont {Poli}, \citenamefont {Prevedelli}, \citenamefont
  {Falke}, \citenamefont {Lisdat}, \citenamefont {Sterr},\ and\ \citenamefont
  {Tino}}]{Schioppo2012}%
  \BibitemOpen
  \bibfield  {author} {\bibinfo {author} {\bibfnamefont {M.}~\bibnamefont
  {Schioppo}}, \bibinfo {author} {\bibfnamefont {N.}~\bibnamefont {Poli}},
  \bibinfo {author} {\bibfnamefont {M.}~\bibnamefont {Prevedelli}}, \bibinfo
  {author} {\bibfnamefont {S.}~\bibnamefont {Falke}}, \bibinfo {author}
  {\bibfnamefont {C.}~\bibnamefont {Lisdat}}, \bibinfo {author} {\bibfnamefont
  {U.}~\bibnamefont {Sterr}},\ and\ \bibinfo {author} {\bibfnamefont {G.~M.}\
  \bibnamefont {Tino}},\ }\href {https://doi.org/10.1063/1.4756936} {\bibfield
  {journal} {\bibinfo  {journal} {Rev. Sci. Instrum.}\ }\textbf {\bibinfo
  {volume} {83}},\ \bibinfo {pages} {103101} (\bibinfo {year}
  {2012})}\BibitemShut {NoStop}%
\bibitem [{\citenamefont {Li}\ \emph {et~al.}(2019)\citenamefont {Li},
  \citenamefont {Chai}, \citenamefont {Wei}, \citenamefont {Yang},
  \citenamefont {Daruwalla}, \citenamefont {Ayaza},\ and\ \citenamefont
  {Raman}}]{Li2019}%
  \BibitemOpen
  \bibfield  {author} {\bibinfo {author} {\bibfnamefont {C.}~\bibnamefont
  {Li}}, \bibinfo {author} {\bibfnamefont {X.}~\bibnamefont {Chai}}, \bibinfo
  {author} {\bibfnamefont {B.}~\bibnamefont {Wei}}, \bibinfo {author}
  {\bibfnamefont {J.}~\bibnamefont {Yang}}, \bibinfo {author} {\bibfnamefont
  {A.}~\bibnamefont {Daruwalla}}, \bibinfo {author} {\bibfnamefont
  {F.}~\bibnamefont {Ayaza}},\ and\ \bibinfo {author} {\bibfnamefont
  {C.}~\bibnamefont {Raman}},\ }\href
  {https://doi.org/10.1038/s41467-019-09647-3} {\bibfield  {journal} {\bibinfo
  {journal} {Nat. Commun.}\ }\textbf {\bibinfo {volume} {10}},\ \bibinfo
  {pages} {1831} (\bibinfo {year} {2019})}\BibitemShut {NoStop}%
\bibitem [{\citenamefont {Ovchinnikov}(2007)}]{Ovchinnikov2007}%
  \BibitemOpen
  \bibfield  {author} {\bibinfo {author} {\bibfnamefont {Y.}~\bibnamefont
  {Ovchinnikov}},\ }\href {https://doi.org/10.1016/j.optcom.2007.04.048}
  {\bibfield  {journal} {\bibinfo  {journal} {Opt. Commun.}\ }\textbf {\bibinfo
  {volume} {276}},\ \bibinfo {pages} {261} (\bibinfo {year}
  {2007})}\BibitemShut {NoStop}%
\bibitem [{\citenamefont {Phillips}\ and\ \citenamefont
  {Metcalf}(1982)}]{Phillips1982}%
  \BibitemOpen
  \bibfield  {author} {\bibinfo {author} {\bibfnamefont {W.~D.}\ \bibnamefont
  {Phillips}}\ and\ \bibinfo {author} {\bibfnamefont {H.~J.}\ \bibnamefont
  {Metcalf}},\ }\href {https://doi.org/10.1103/PhysRevLett.48.596} {\bibfield
  {journal} {\bibinfo  {journal} {Phys. Rev. Lett.}\ }\textbf {\bibinfo
  {volume} {48}},\ \bibinfo {pages} {596} (\bibinfo {year} {1982})}\BibitemShut
  {NoStop}%
\bibitem [{\citenamefont {Dareau}(2015)}]{Dareau2015}%
  \BibitemOpen
  \bibfield  {author} {\bibinfo {author} {\bibfnamefont {A.}~\bibnamefont
  {Dareau}},\ }\emph {\bibinfo {title} {Manipulation cohérente d'un condensat
  de {B}ose-{E}instein d'ytterbium sur la transition d'horloge: de la
  spectroscopie au magnétisme artificiel}},\ \href@noop {} {Ph.D. thesis},\
  \bibinfo  {school} {École Normale Supérieure} (\bibinfo {year} {2015}),\
  \bibinfo {note}
  {\url{https://tel.archives-ouvertes.fr/tel-01194429}}\BibitemShut {NoStop}%
\bibitem [{\citenamefont {Hopkins}\ \emph {et~al.}(2016)\citenamefont
  {Hopkins}, \citenamefont {Butler}, \citenamefont {Guttridge}, \citenamefont
  {Kemp}, \citenamefont {Freytag}, \citenamefont {Hinds}, \citenamefont
  {Tarbutt},\ and\ \citenamefont {Cornish}}]{Hopkins2016}%
  \BibitemOpen
  \bibfield  {author} {\bibinfo {author} {\bibfnamefont {S.}~\bibnamefont
  {Hopkins}}, \bibinfo {author} {\bibfnamefont {K.}~\bibnamefont {Butler}},
  \bibinfo {author} {\bibfnamefont {A.}~\bibnamefont {Guttridge}}, \bibinfo
  {author} {\bibfnamefont {S.}~\bibnamefont {Kemp}}, \bibinfo {author}
  {\bibfnamefont {R.}~\bibnamefont {Freytag}}, \bibinfo {author} {\bibfnamefont
  {E.}~\bibnamefont {Hinds}}, \bibinfo {author} {\bibfnamefont
  {M.}~\bibnamefont {Tarbutt}},\ and\ \bibinfo {author} {\bibfnamefont
  {S.}~\bibnamefont {Cornish}},\ }\href {https://doi.org/10.1063/1.4945795}
  {\bibfield  {journal} {\bibinfo  {journal} {Rev. Sci. Instrum.}\ }\textbf
  {\bibinfo {volume} {87}},\ \bibinfo {pages} {043109} (\bibinfo {year}
  {2016})}\BibitemShut {NoStop}%
\bibitem [{\citenamefont {Freytag}(2015)}]{Freytag2015}%
  \BibitemOpen
  \bibfield  {author} {\bibinfo {author} {\bibfnamefont {R.}~\bibnamefont
  {Freytag}},\ }\emph {\bibinfo {title} {Simultaneous magneto-optical trapping
  of ytterbium and caesium}},\ \href {https://doi.org/10.25560/26874} {Ph.D.
  thesis},\ \bibinfo  {school} {Imperial College London} (\bibinfo {year}
  {2015})\BibitemShut {NoStop}%
\bibitem [{\citenamefont {Halbach}(1980)}]{Halbach1980}%
  \BibitemOpen
  \bibfield  {author} {\bibinfo {author} {\bibfnamefont {K.}~\bibnamefont
  {Halbach}},\ }\href {https://doi.org/10.1016/0029-554X(80)90094-4} {\bibfield
   {journal} {\bibinfo  {journal} {Nucl. Instrum. Methods}\ }\textbf {\bibinfo
  {volume} {169}},\ \bibinfo {pages} {1} (\bibinfo {year} {1980})}\BibitemShut
  {NoStop}%
\bibitem [{\citenamefont {Ben~Ali}\ \emph {et~al.}(2017)\citenamefont
  {Ben~Ali}, \citenamefont {Badr}, \citenamefont {Br{\'e}zillon}, \citenamefont
  {Dubessy}, \citenamefont {Perrin},\ and\ \citenamefont
  {Perrin}}]{BenAli2017}%
  \BibitemOpen
  \bibfield  {author} {\bibinfo {author} {\bibfnamefont {D.}~\bibnamefont
  {Ben~Ali}}, \bibinfo {author} {\bibfnamefont {T.}~\bibnamefont {Badr}},
  \bibinfo {author} {\bibfnamefont {T.}~\bibnamefont {Br{\'e}zillon}}, \bibinfo
  {author} {\bibfnamefont {R.}~\bibnamefont {Dubessy}}, \bibinfo {author}
  {\bibfnamefont {H.}~\bibnamefont {Perrin}},\ and\ \bibinfo {author}
  {\bibfnamefont {A.}~\bibnamefont {Perrin}},\ }\href
  {https://doi.org/10.1088/1361-6455/aa5a6a} {\bibfield  {journal} {\bibinfo
  {journal} {J. Phys. B: At. Mol. and Opt. Phys.}\ }\textbf {\bibinfo {volume}
  {50}},\ \bibinfo {pages} {055008} (\bibinfo {year} {2017})}\BibitemShut
  {NoStop}%
\bibitem [{\citenamefont {Cheiney}\ \emph {et~al.}(2011)\citenamefont
  {Cheiney}, \citenamefont {Carraz}, \citenamefont {Bartoszek-Bober},
  \citenamefont {Faure}, \citenamefont {Vermersch}, \citenamefont {Fabre},
  \citenamefont {Gattobigio}, \citenamefont {Lahaye}, \citenamefont
  {Gu{\'e}ry-Odelin},\ and\ \citenamefont {Mathevet}}]{Cheiney2011}%
  \BibitemOpen
  \bibfield  {author} {\bibinfo {author} {\bibfnamefont {P.}~\bibnamefont
  {Cheiney}}, \bibinfo {author} {\bibfnamefont {O.}~\bibnamefont {Carraz}},
  \bibinfo {author} {\bibfnamefont {D.}~\bibnamefont {Bartoszek-Bober}},
  \bibinfo {author} {\bibfnamefont {S.}~\bibnamefont {Faure}}, \bibinfo
  {author} {\bibfnamefont {F.}~\bibnamefont {Vermersch}}, \bibinfo {author}
  {\bibfnamefont {C.}~\bibnamefont {Fabre}}, \bibinfo {author} {\bibfnamefont
  {G.}~\bibnamefont {Gattobigio}}, \bibinfo {author} {\bibfnamefont
  {T.}~\bibnamefont {Lahaye}}, \bibinfo {author} {\bibfnamefont
  {D.}~\bibnamefont {Gu{\'e}ry-Odelin}},\ and\ \bibinfo {author} {\bibfnamefont
  {R.}~\bibnamefont {Mathevet}},\ }\href {https://doi.org/10.1063/1.3600897}
  {\bibfield  {journal} {\bibinfo  {journal} {Rev. Sci. Instrum.}\ }\textbf
  {\bibinfo {volume} {82}},\ \bibinfo {pages} {063115} (\bibinfo {year}
  {2011})}\BibitemShut {NoStop}%
\bibitem [{Note1()}]{Note1}%
  \BibitemOpen
  \bibinfo {note} {HKCM Engineering, part no. Q128x06x06Zn-30SH}\BibitemShut
  {NoStop}%
\bibitem [{Note2()}]{Note2}%
  \BibitemOpen
  \bibinfo {note} {HKCM Engineering, part no. Q25x04x04Zn-35H}\BibitemShut
  {NoStop}%
\bibitem [{Note3()}]{Note3}%
  \BibitemOpen
  \bibinfo {note} {F.W. Bell, Model 7030}\BibitemShut {NoStop}%
\bibitem [{\citenamefont {Stellmer}(2013)}]{Stellmer2013}%
  \BibitemOpen
  \bibfield  {author} {\bibinfo {author} {\bibfnamefont {S.}~\bibnamefont
  {Stellmer}},\ }\emph {\bibinfo {title} {Degenerate quantum gases of
  strontium}},\ \href@noop {} {Ph.D. thesis},\ \bibinfo  {school} {University
  of Innsbruck} (\bibinfo {year} {2013}),\ \bibinfo {note}
  {\url{http://www.ultracold.at/theses/2013-stellmer.pdf}}\BibitemShut
  {NoStop}%
\bibitem [{\citenamefont {Huckans}\ \emph {et~al.}(2018)\citenamefont
  {Huckans}, \citenamefont {Dubosclard}, \citenamefont {Mar\'echal},
  \citenamefont {Gorceix}, \citenamefont {Laburthe-Tolra},\ and\ \citenamefont
  {Robert-de Saint-Vincent}}]{Huckans2018}%
  \BibitemOpen
  \bibfield  {author} {\bibinfo {author} {\bibfnamefont {J.}~\bibnamefont
  {Huckans}}, \bibinfo {author} {\bibfnamefont {W.}~\bibnamefont {Dubosclard}},
  \bibinfo {author} {\bibfnamefont {E.}~\bibnamefont {Mar\'echal}}, \bibinfo
  {author} {\bibfnamefont {B.}~\bibnamefont {Gorceix}}, \bibinfo {author}
  {\bibfnamefont {B.~O.}\ \bibnamefont {Laburthe-Tolra}},\ and\ \bibinfo
  {author} {\bibfnamefont {M.}~\bibnamefont {Robert-de Saint-Vincent}},\ }\href
  {https://arxiv.org/abs/1802.08499} {\bibinfo {title} {Note on the reflectance
  of mirrors exposed to a strontium beam}} (\bibinfo {year} {2018}),\ \Eprint
  {https://arxiv.org/abs/1802.08499} {arXiv:1802.08499 [physics.atom-ph]}
  \BibitemShut {NoStop}%
\bibitem [{Note4()}]{Note4}%
  \BibitemOpen
  \bibinfo {note} {HKCM Engineering Q80x08x06Zn-35H}\BibitemShut {NoStop}%
\bibitem [{\citenamefont {Tiecke}\ \emph {et~al.}(2009)\citenamefont {Tiecke},
  \citenamefont {A.}, \citenamefont {Ludewig},\ and\ \citenamefont
  {Walraven}}]{Tiecke2009}%
  \BibitemOpen
  \bibfield  {author} {\bibinfo {author} {\bibfnamefont {G.}~\bibnamefont
  {Tiecke}, \bibfnamefont {T.~G.}}, \bibinfo {author} {\bibnamefont {A.}},
  \bibinfo {author} {\bibfnamefont {A.}~\bibnamefont {Ludewig}},\ and\ \bibinfo
  {author} {\bibfnamefont {J.~T.~M.}\ \bibnamefont {Walraven}},\ }\href
  {https://doi.org/10.1103/PhysRevA.800.013409} {\bibfield  {journal} {\bibinfo
   {journal} {Phys. Rev. A}\ }\textbf {\bibinfo {volume} {80}},\ \bibinfo
  {pages} {013409} (\bibinfo {year} {2009})}\BibitemShut {NoStop}%
\bibitem [{\citenamefont {Cho}\ \emph {et~al.}(2012)\citenamefont {Cho},
  \citenamefont {Lee}, \citenamefont {Lee}, \citenamefont {Ahn}, \citenamefont
  {Lee}, \citenamefont {Yu}, \citenamefont {Lee},\ and\ \citenamefont
  {Park}}]{Cho2012}%
  \BibitemOpen
  \bibfield  {author} {\bibinfo {author} {\bibfnamefont {J.~W.}\ \bibnamefont
  {Cho}}, \bibinfo {author} {\bibfnamefont {H.}~\bibnamefont {Lee}}, \bibinfo
  {author} {\bibfnamefont {S.}~\bibnamefont {Lee}}, \bibinfo {author}
  {\bibfnamefont {J.}~\bibnamefont {Ahn}}, \bibinfo {author} {\bibfnamefont
  {W.}~\bibnamefont {Lee}}, \bibinfo {author} {\bibfnamefont {D.}~\bibnamefont
  {Yu}}, \bibinfo {author} {\bibfnamefont {S.~K.}\ \bibnamefont {Lee}},\ and\
  \bibinfo {author} {\bibfnamefont {C.~Y.}\ \bibnamefont {Park}},\ }\href
  {https://doi.org/10.1103/PhysRevA.85.035401} {\bibfield  {journal} {\bibinfo
  {journal} {Phys. Rev. A}\ }\textbf {\bibinfo {volume} {85}},\ \bibinfo
  {pages} {035401} (\bibinfo {year} {2012})}\BibitemShut {NoStop}%
\bibitem [{\citenamefont {Guttridge}\ \emph {et~al.}(2016)\citenamefont
  {Guttridge}, \citenamefont {Hopkins}, \citenamefont {L.}, \citenamefont
  {Boddy}, \citenamefont {Freytag}, \citenamefont {Jones}, \citenamefont {R.},
  \citenamefont {Hinds},\ and\ \citenamefont {L.}}]{Guttridge2016}%
  \BibitemOpen
  \bibfield  {author} {\bibinfo {author} {\bibfnamefont {A.}~\bibnamefont
  {Guttridge}}, \bibinfo {author} {\bibfnamefont {S.~A.}\ \bibnamefont
  {Hopkins}}, \bibinfo {author} {\bibfnamefont {K.~S.}\ \bibnamefont {L.}},
  \bibinfo {author} {\bibfnamefont {D.}~\bibnamefont {Boddy}}, \bibinfo
  {author} {\bibfnamefont {R.}~\bibnamefont {Freytag}}, \bibinfo {author}
  {\bibfnamefont {M.~P.~A.}\ \bibnamefont {Jones}}, \bibinfo {author}
  {\bibfnamefont {T.~M.}\ \bibnamefont {R.}}, \bibinfo {author} {\bibfnamefont
  {E.~A.}\ \bibnamefont {Hinds}},\ and\ \bibinfo {author} {\bibfnamefont
  {C.~S.}\ \bibnamefont {L.}},\ }\href
  {https://doi.org/10.1088/0953-4075/49/14/145006} {\bibfield  {journal}
  {\bibinfo  {journal} {J. Phys. B}\ }\textbf {\bibinfo {volume} {49}},\
  \bibinfo {pages} {145006} (\bibinfo {year} {2016})}\BibitemShut {NoStop}%
\bibitem [{\citenamefont {Lunden}\ \emph {et~al.}(2020)\citenamefont {Lunden},
  \citenamefont {Du}, \citenamefont {Cantara}, \citenamefont {Barral},
  \citenamefont {Jamison},\ and\ \citenamefont {Ketterle}}]{Lunden2020}%
  \BibitemOpen
  \bibfield  {author} {\bibinfo {author} {\bibfnamefont {W.}~\bibnamefont
  {Lunden}}, \bibinfo {author} {\bibfnamefont {L.}~\bibnamefont {Du}}, \bibinfo
  {author} {\bibfnamefont {M.}~\bibnamefont {Cantara}}, \bibinfo {author}
  {\bibfnamefont {P.}~\bibnamefont {Barral}}, \bibinfo {author} {\bibfnamefont
  {A.~O.}\ \bibnamefont {Jamison}},\ and\ \bibinfo {author} {\bibfnamefont
  {W.}~\bibnamefont {Ketterle}},\ }\href
  {https://doi.org/10.1103/PhysRevA.101.063403} {\bibfield  {journal} {\bibinfo
   {journal} {Phys. Rev. A}\ }\textbf {\bibinfo {volume} {101}},\ \bibinfo
  {pages} {063403} (\bibinfo {year} {2020})}\BibitemShut {NoStop}%
\end{thebibliography}%
